\documentclass[12pt,preprint]{aastex}

\begin{document}
\title{Three-Dimensional Simulations of a Starburst-Driven Galactic Wind}

\shorttitle{3D Simulations of a Starburst Wind}

\author{Jackie L. Cooper, Geoffrey  V. Bicknell, Ralph  S. Sutherland,}
\affil{Research School of Astronomy and Astrophysics,\\ 
  The Australian National University, Cotter Road,\\
  Weston Creek, ACT 2611, Australia\\}
\email{jcooper@mso.anu.edu.au}
\and
\author{Joss Bland-Hawthorn$^{1,2}$}
\affil{$^1$ Institute of Astronomy, School of Physics,\\ University of Sydney, NSW 2006, Australia\\ 
$^2$ Anglo-Australian Observatory, P.O. Box 296,\\ 
Epping, NSW 2121, Australia\\}

\begin{abstract}
We have performed a series of three-dimensional simulations of a
starburst-driven wind in an inhomogeneous interstellar
medium. The introduction of an inhomogeneous disk leads to differences in the
formation of a wind, most noticeably the absence of the ``blow-out'' effect
seen in homogeneous models. A wind forms from a series of small bubbles that
propagate into the tenuous gas between dense clouds in the disk. These bubbles
merge and follow the path of least resistance out of the disk, before flowing
freely into the halo. Filaments are formed from disk gas that is broken up and accelerated into
the outflow. These filaments are distributed throughout a biconical structure
within a more spherically distributed hot wind. The distribution of the
inhomogeneous interstellar medium in the disk is important in determining the
morphology of this wind, as well as the distribution of the
filaments. While higher resolution simulations are required in order to
ascertain the importance of mixing processes, we find that soft X-ray emission
arises from gas that has been mass-loaded from clouds in the disk, as well as
from bow shocks upstream of clouds, driven into the flow by the ram pressure of the wind, and the
interaction between these shocks.
\end{abstract}

\section{INTRODUCTION}

In 1963, \citeauthor{LS1963} first detected an outflow of gas along
the minor axis of M82. \citet{CC1985} proposed a model in which a galactic scale outflow
could be powered by the combined kinetic energy from supernovae. Starburst galaxies,
with their characteristically high star formation rates, provide the perfect
environments for these winds to develop. Indeed, galactic winds are ubiquitous in starburst
galaxies, having been observed in many nearby galaxies and inferred in
galaxies at  high-redshifts \citep[see][~and references therein]{VCB2005}  

The best studied galactic wind is the outflow in M82, which is clearly visible in the
light of H$\alpha$, displaying a vast filamentary system extending several kpc along
the minor axis of the galaxy \citep{SB1998}. These filaments lie on the surface of a
mostly hollow structure and rotate in the same direction as the disk \citep{Greve2004}. 
As with other galactic winds \citep[e.g. NGC 253:][]{SDW2003},
the wind in M82 is asymmetric, with the northern outflow more chaotic
than the southern outflow. The filaments can be traced to the nuclear
region and display both shell and loop-like structures
\citep{Oetal2002}. The formation of these filaments is currently not well understood, but they are
thought to be either disk or halo gas that has been entrained into
the outflow.

The morphology of galactic winds can vary. Outflows often
display asymmetries, varying degrees of collimation and may be tilted with respect to the
minor axis. While many outflows are limb-brightened \citep[e.g. NGC
  3079;][]{Vetal1994}, the optical filaments can also fill
the volume rather than  remain confined to the surface of the biconical outflow
\citep{VB1997}. The host galaxy itself plays an important role in determining the
morphology of a wind, with its size and structure affecting the degree of collimation
\citep{SS2000} and expansion of the outflow \citep{SetalB2004,Getal2005,Martin2005}.  

Recent Chandra observations have revealed increasing detail in the X-ray
emission from galactic winds. One of the most striking results of these
observations is the close spatial relationship with the H$\alpha$
emitting gas
\citep[e.g.][]{Setal2000,Setal2002,SetalA2004,CBV2002,MKH2002,Getal2005,OWB2005b},
suggesting a close physical connection. Thus, a successful model of a galactic
wind needs to explain this
relationship. \citet{Setal2002} provide a summary of several theories for the
origin of the X-ray emission that could explain this correlation. These
mechanisms involve shocked disk or halo gas that has been swept up
into the wind, in the form of dense clouds or shells.      

Over the past few decades, numerous simulations have been made of starburst-driven winds
\citep{TI1988,TB1993,Setal1994,Setal1996,DB1999,TM1998,SS2000,TSM2003}.
\citet{Setal1994} performed two-dimensional, axisymmetric simulations of a
galactic wind in an isothermal ISM, with varying densities and temperatures. They
concluded that the H$\alpha$ filaments form from disk gas that has been entrained
into the flow and that the X-ray emission most likely arises from shocked disk
and halo gas. More recently, \citet{SS2000} performed a series of simulations,
focusing on the energetics and X-ray emission from the wind. As with
\citet{Setal1994}, their simulations were two-dimensional and axisymmetric with
an isothermal ISM. They found that a large fraction of the soft X-ray emission
in their model comes from shock-heated ambient gas and from the interfaces
between cool dense and hot tenuous gas. While these simulations provide some
insight into the origin of the X-ray and H$\alpha$ emission, the homogeneous
nature of these  models and their symmetry renders them incapable of
forming significant filamentary structures, limiting their ability to
constrain the emission processes.      

In order to improve upon previous models and to gain a better understanding of
the origin of the H$\alpha$ filaments and X-ray emission, we have performed a
series of three-dimensional simulations of a galactic wind in an inhomogeneously
distributed interstellar medium (ISM). The introduction of inhomogeneity is important as
the interstellar medium in a galaxy disk is highly complex in all its phases
\citep[see for example][~and references therein]{EE2001}.  Inhomogeneity is also crucial 
in the development of a wind, as energy from massive stars formed in dense
molecular clouds in the starburst region may be radiated away before a wind
could form. A wind is more likely to develop from the kinetic energy from
stellar winds adjacent to the diffuse gas surrounding the clouds.

The inhomogeneous structure of the ISM is also likely to affect the
distribution of filaments throughout the wind,
producing asymmetric and tilted outflows. It is likely that the
size and strength of the starburst itself plays an important role in determining the
morphology. Many starburst galaxies, such as M82 and NGC 3079, contain
circumnuclear starbursts, with their resultant outflows being strong and
violent \citep{SB1998,Vetal1994}. Other starbursts are weaker and have less
prominent outflows. An example is NGC 4631, which is currently undergoing a
disk-wide starburst \citep{SetalA2004}.

\citet{TSM2003} investigated the formation of the emission line filaments by modeling the
formation of a wind from several super star clusters. They proposed that kiloparsec long filaments
are formed from stationary and oblique shocks. In this paper we present a
different model, which follows a similar approach to that of \citet{SS2000},
but introduces an inhomogeneous disk. We follow the
evolution of a starburst-driven wind in different ISM
conditions and discuss the effect of the inhomogeneity of the disk on the
morphology of the wind. We consider the morphology of the H$\alpha$ emitting
filaments separately and investigate their origin. Finally, the luminosity of the soft and
hard X-ray emitting gas is calculated and we suggest an origin for the soft
X-ray emission. 

\section{NUMERICAL MODEL}

\subsection{Description of the Code}

The simulations were performed using a PPMLR code (Piecewise Parabolic
Method with a Lagrangian Remap), which is based on the method
described by \citet{CW1984}. The code has been extensively modified
\citep[see, for example,~][]{SBB2003,SBD2003}
from the original VH-1 code \citep{Blondin1995}. It is
a multi-dimensional hydrodynamics code, optimized for use on multiple
processors of the SGI Altix computer operated by the Australian
Partnership for Advanced Computing (APAC). Thermal cooling has been
incorporated, based upon the output from the MAPPINGS III code
\citep[see][]{SD1993,SBB2003,Setal2005}, enabling the realistic evolution of a radiatively cooling
gas. The simulations discussed in this paper
are three-dimensional and utilize cartesian (x,y,z) 
coordinates. In each cell of the computational grid, the density,
temperature, velocity, emissivity and a disk gas tracer are recorded at intervals of 0.01 Myr.

\subsubsection{The Gravitational Potential}

Following \citet{SS2000}, the gravitational potential used in
these simulations consists of a stellar spheroid and a
disk. Let $R$ = $\sqrt{r^2 + z^2}$ be the radius of the stellar spheroid, $r_0$ the core radius,
$M_{\rm{ss}}$ its mass, $M_{\rm{disk}}$ 
the mass of the disk, $a$  its radial scale length, and
$b$ its vertical scale length. The potential
$\Phi_{\rm{ss}}$ of the stellar spheroid is described by an analytic King
model (eq. [\ref{ss}]) and the disk potential $\Phi_{\rm{disk}}$ by a
\citet{MN1975} model (eq. [\ref{disk}]). The total gravitational potential is
then the sum of the two components
($\Phi_{\rm{tot}} = \Phi_{\rm{ss}} + \Phi_{\rm{disk}}$), where

\begin{equation}\label{ss}
\Phi_{\rm{ss}}(R) =
-\frac{GM_{\rm{ss}}}{r_0}\left\{\frac{\ln\left[(R/r_0)+\sqrt{1+(R/r_0)^2}\right]}
{(R/r_0)}\right\}
\end{equation}

\begin{equation}\label{disk}
\Phi_{\rm{disk}}(r,z) = -\frac{GM_{\rm{disk}}}{\sqrt{r^2 +
    (a+\sqrt{z^2 +b^2})^2}}
\end{equation}

Whilst these simulations are intended to be applicable to the general class of disk galaxies, 
we used parameters which are based on the iconic galaxy, M82. Note also that simulations 
including cooling admit a one-parameter scaling which is described in \citet{SB2007}. 
Within limits the simulations may be scaled to smaller or larger galaxies.

\begin{figure}[t]
\epsscale{0.75}
\plotone{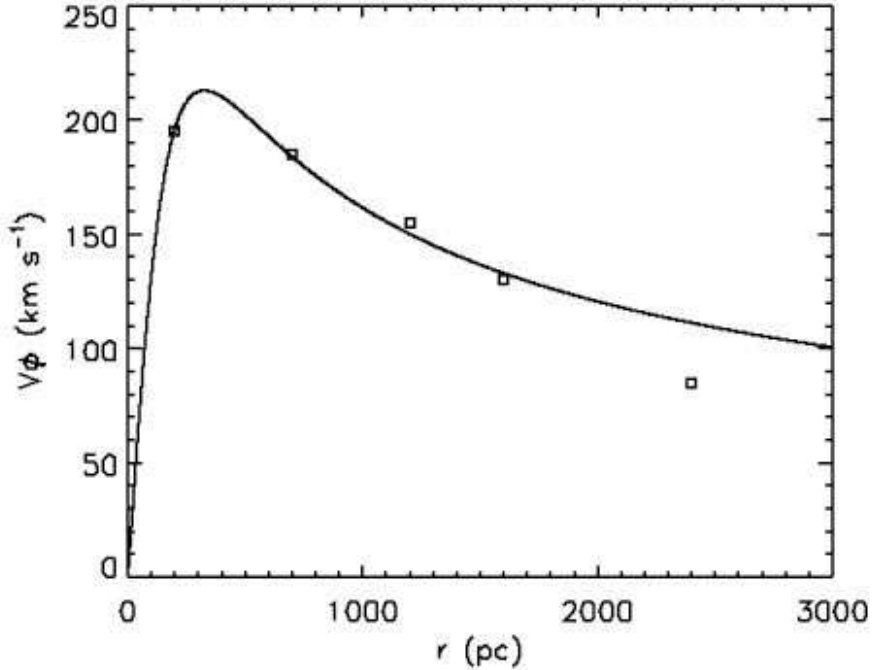}
\caption{Fit to the CO rotation curve of M82 \cite[empty squares:][]{Sofue1998}}\label{fig1}
\end{figure}

We adopted parameters for the above potential by approximately fitting the
rotation curve of M82 (Figure \ref{fig1}); the parameters are
summarized in Table \ref{tab_param}. This produces a good
fit at the smaller radii used in these simulations. We neglect the
contribution of a dark matter halo, since our model only extends to a
radius of less than 1 kpc. This is justified because, for example, in the
Galaxy where the contribution
of dark matter is well constrained at all radii, it is
now well established that baryonic matter dominates the
potential within the Solar Circle \citep[see][~for review]{Binney2005}. 


\begin{figure}[t]
\includegraphics[scale=0.4]{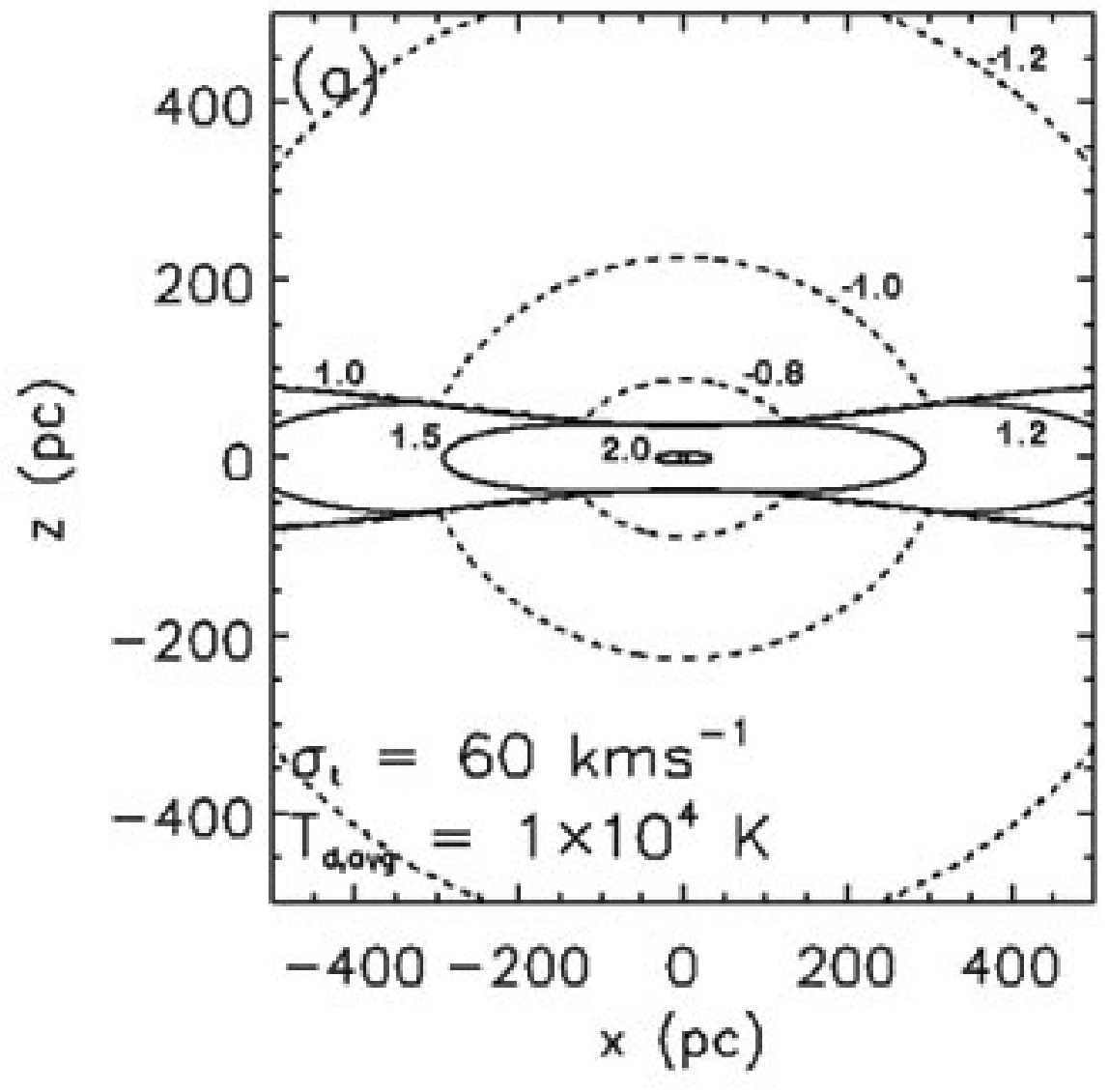}
\includegraphics[scale=0.4]{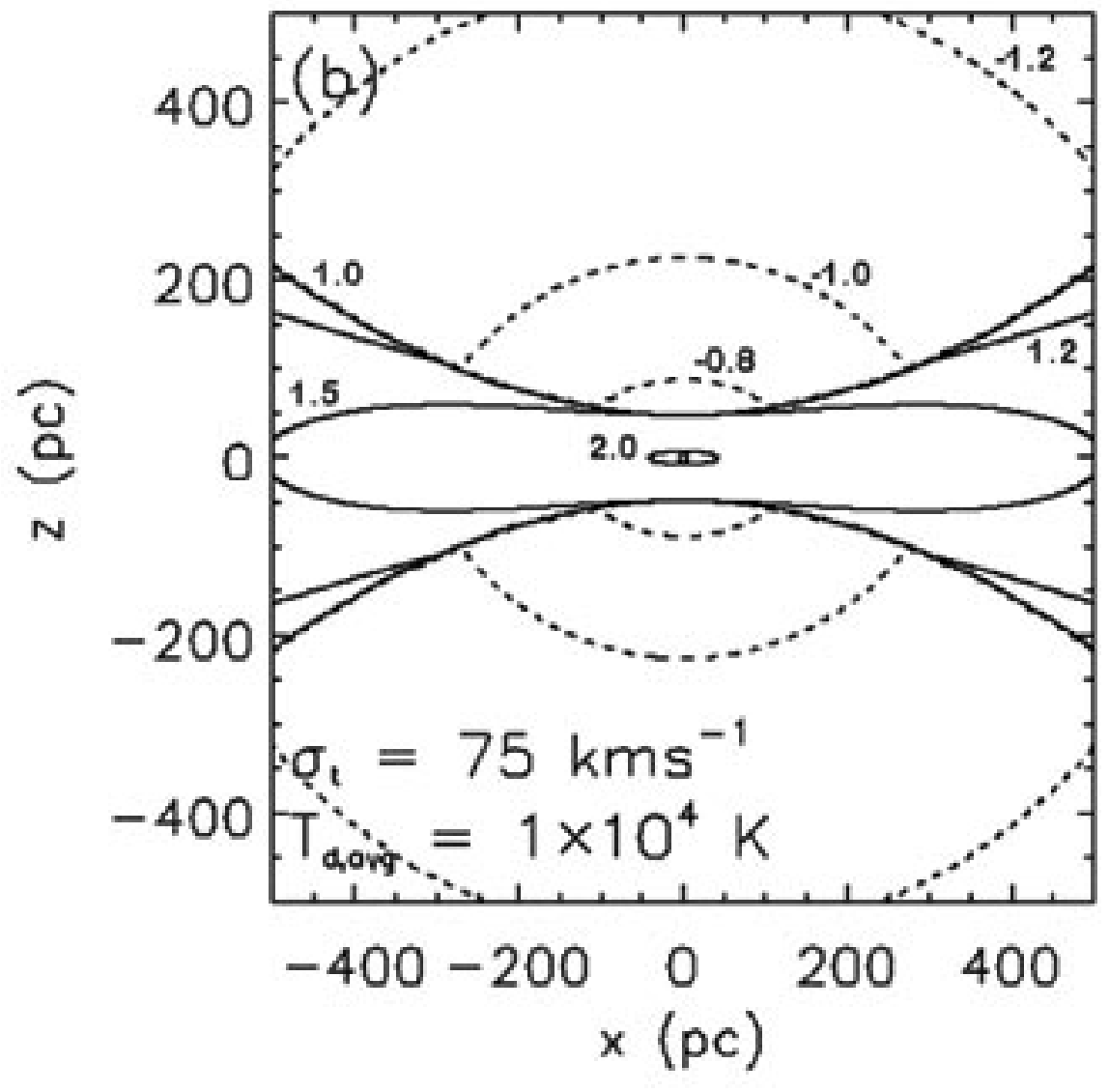}
\includegraphics[scale=0.4]{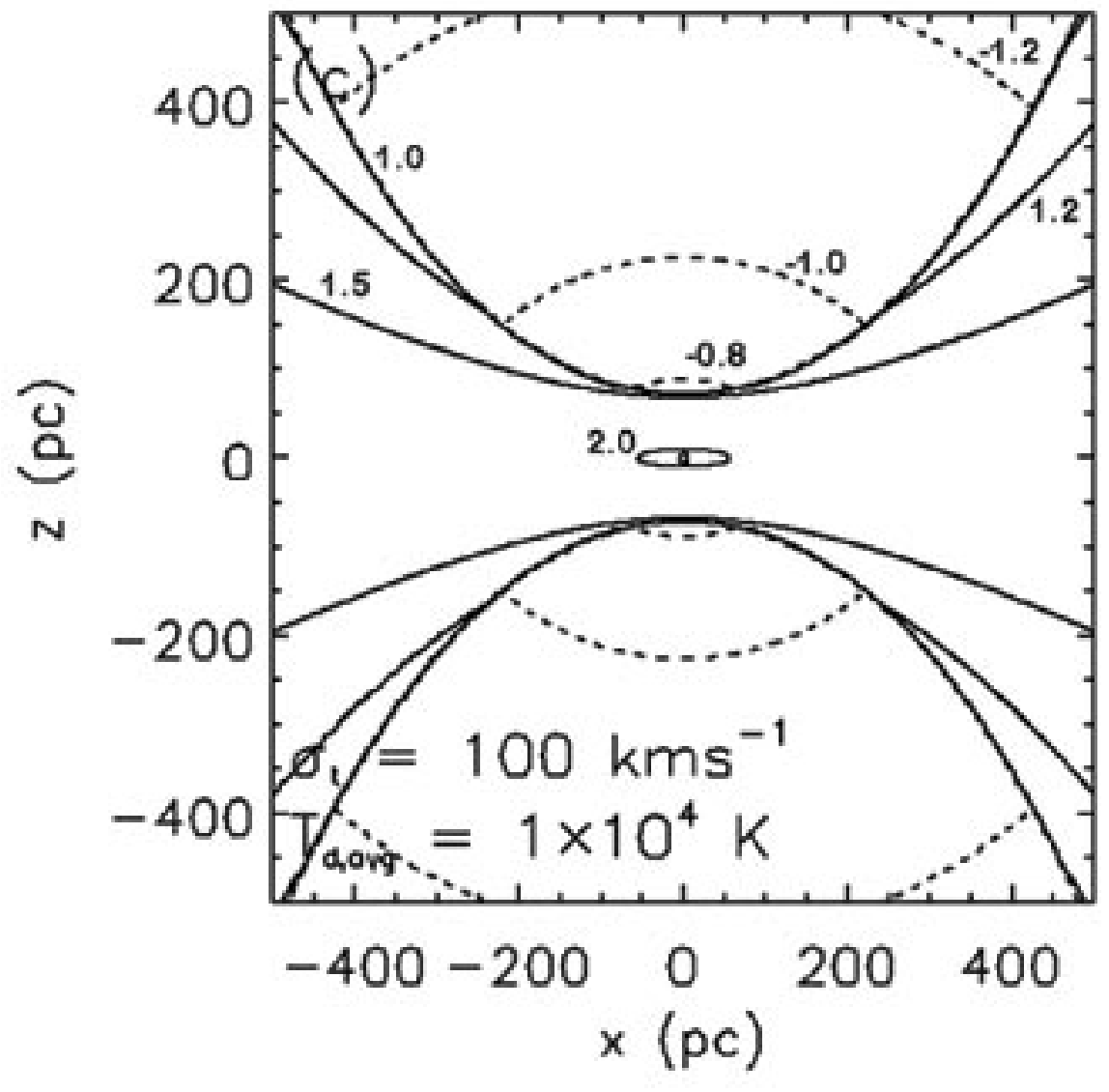}\\
\includegraphics[scale=0.4]{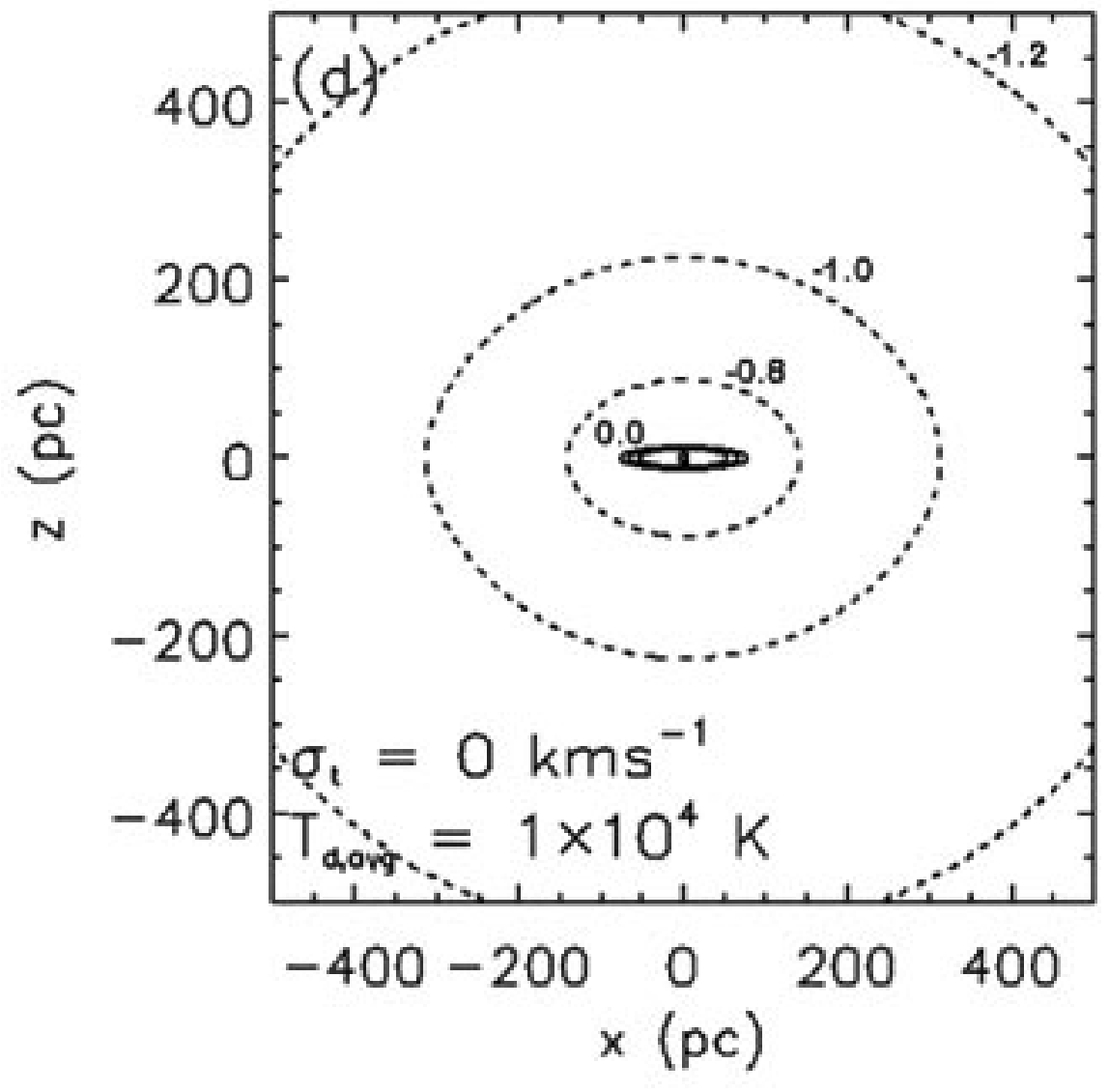}
\includegraphics[scale=0.4]{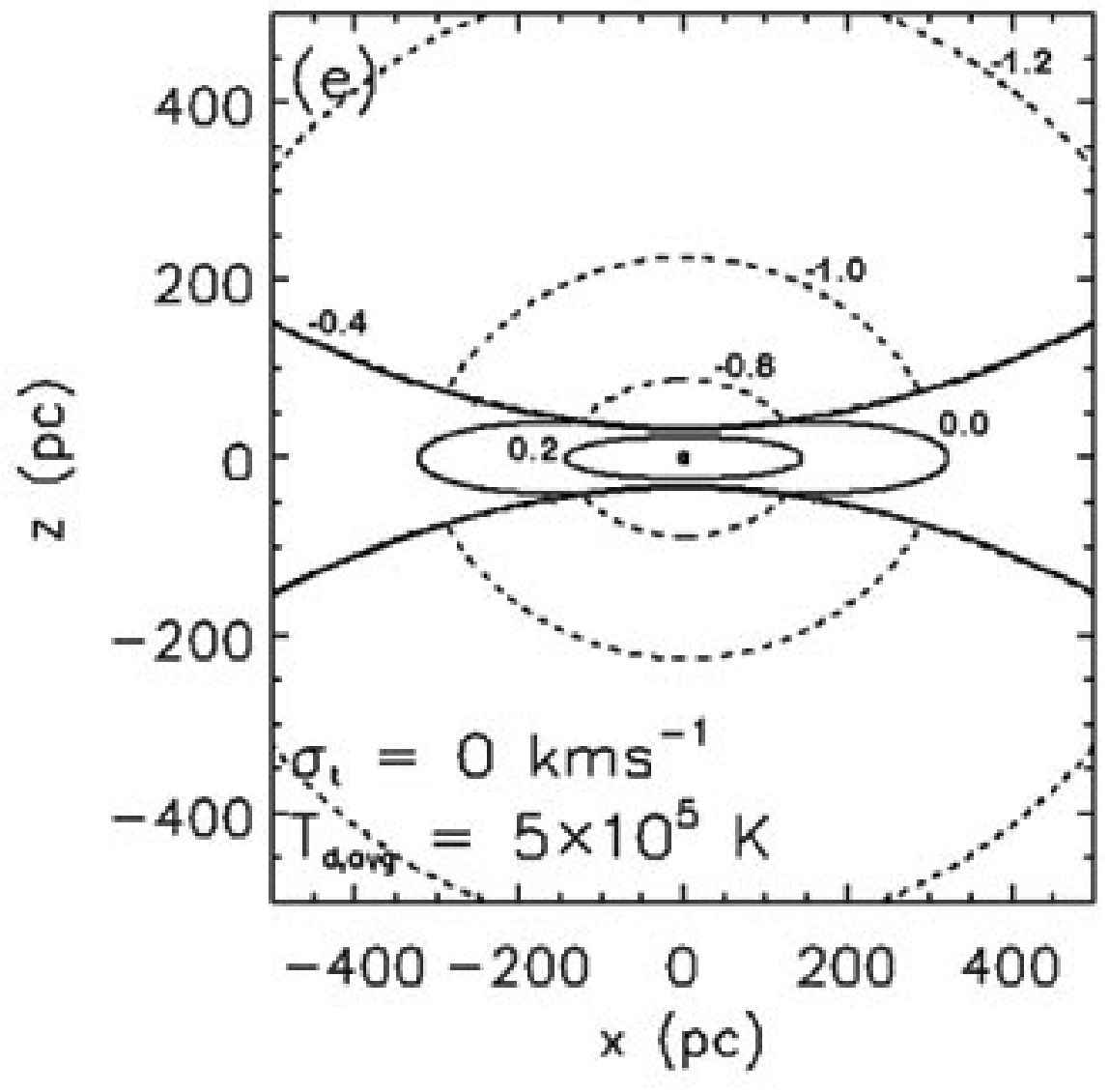}
\includegraphics[scale=0.4]{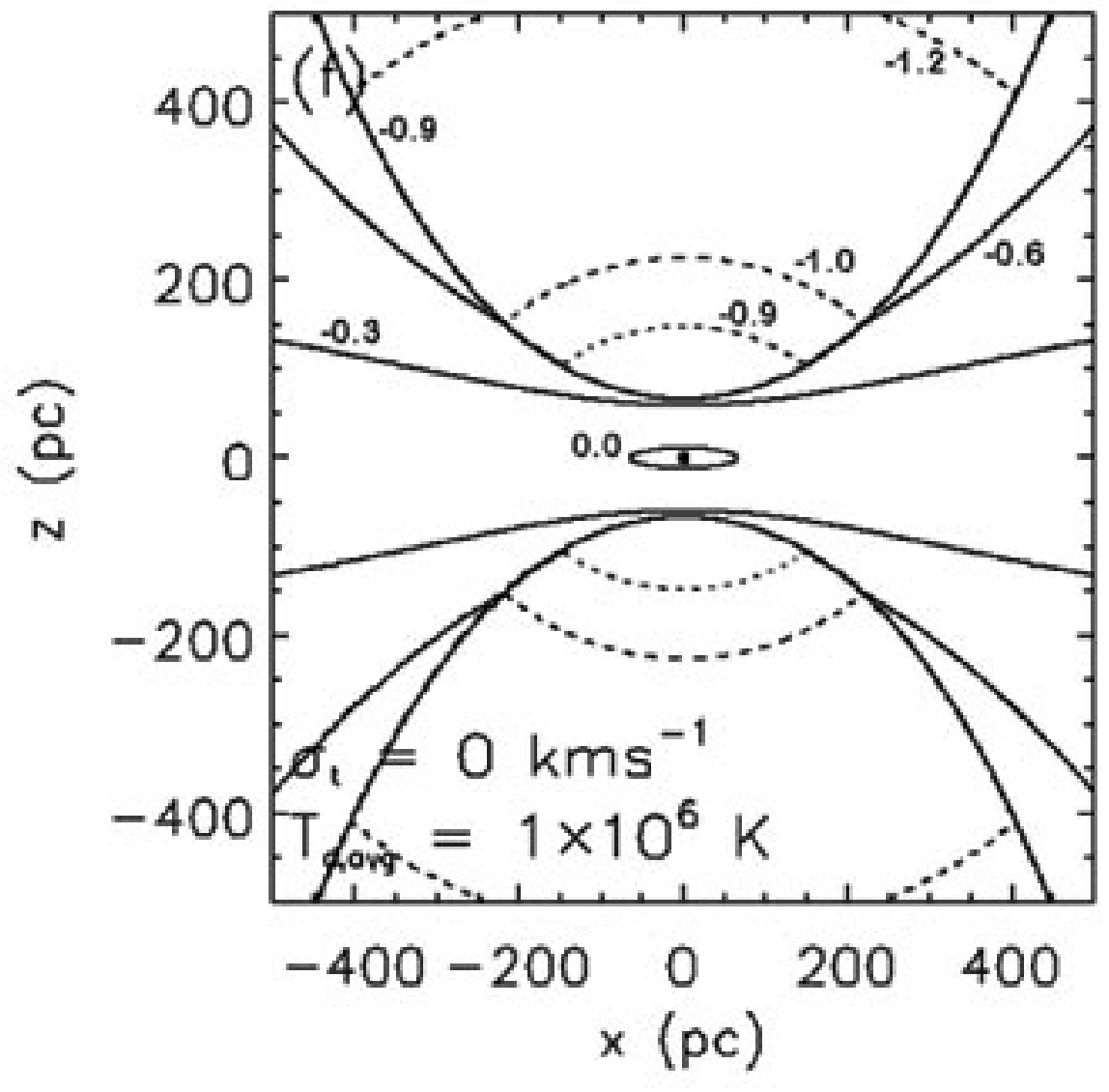}
\caption{Log-density contours of the disk and halo showing the effect the
  parameters $\sigma_{\rm t}$ (top row)
and $T_{\rm d}$ (bottom row) on the scale height of the disk. The solid contours
represent the disk gas, while the dashed contours represents the halo gas.}\label{cont}
\end{figure}

\begin{deluxetable}{lcc}
\tablewidth{0pt}
\tablecaption{Model Parameters \label{tab_param}}
\tablehead{
\colhead{Parameter} & \colhead{Symbol}  & \colhead{Value}}
\startdata
Stellar spheroid mass & M$_{\rm{ss}}$ & 6.0 $\times$ 10$^{8}$ M$_{\odot}$\\
Disk mass & M$_{\rm{disk}}$ & 6.0 $\times$ 10$^{9}$ M$_{\odot}$\\
Core radius & r$_{0}$ & 350.0 pc\\
Radial scale length & a & 150.0 pc\\
Vertical scale length & b  & 75.0 pc\\ 
Central halo density & n$_{\rm{h}}$  & 0.2 cm$^{-3}$\\
Average disk density & n$_{\rm{d,avg}}$ & 100.0 cm$^{-3}$\\
Halo temperature & T$_{\rm{h}}$ &  5.0 $\times$ 10$^{6}$ K\\
Average disk temperature & T$_{\rm{d,avg}}$ &  1.0 $\times$ 10$^{4}$ K\\
Starburst radius & $r_{\rm{sb}}$ & 150.0 pc\\
Starburst height & $h_{\rm{sb}}$ & 60.0 pc\\
Mass injection rate & $\dot{M}$ & 1.0 M$_{\odot}$~yr$^{-1}$\\
Energy injection rate & $\dot{E}$ & 1.0 $\times$ 10$^{42}$ erg~s$^{-1}$\\
\enddata
\end{deluxetable}


\subsubsection{The Interstellar Medium}

The interstellar medium used in these simulations has two components, a hot
isothermal halo and a turbulent warm inhomogeneous disk. As with \citet{SS2000}, the
density distribution of the halo is homogeneous and is described by equation
(\ref{hot}), where $c_{\rm{s,h}}$ = $\sqrt{kT_{\rm{h}}/{\mu}m}$ is
the isothermal sound speed of the hot gas, and $e_{\rm{h}}$ is the ratio of the azimuthal
velocity to the Keplerian velocity. In order to obtain a non-rotating halo,
which is supported by pressure alone, we adopt $e_{\rm{h}} = 0$.     

\begin{equation}\label{hot}
\frac{\rho_{\rm{halo}}(r,z)}{\rho_{\rm{halo}}(0,0)} =
\exp\left[-\frac{\Phi_{\rm{tot}}(r,z) -
	e_{\rm{h}}^2\Phi_{\rm{tot}}(r,0) -
	(1-e_{\rm{h}}^2)\Phi_{\rm{tot}}(0,0)}{c_{\rm{s,h}}^2}\right]
\end{equation}

We follow the same approach as \citet{SB2007} and introduce an ensemble mean
density distribution which introduces a turbulence parameter. The introduction
of turbulence removes the need to impose artificially high temperatures to
achieve reasonable disk scale heights. We discuss the details of this
distribution below. In order to construct the warm inhomogeneous disk, we
first establish the ensemble mean density
distribution. Let $c_{\rm{s,d}}$ = $\sqrt{kT_{\rm{d}}/{\mu}m}$ be the
sound speed of the warm gas, $\sigma_{\rm t}$ the turbulent velocity
dispersion of the clouds, and
$e_{\rm{d}}$ be the ratio of azimuthal to the Keplerian velocity of the warm
gas. As shown by \citet{SB2007}, the parameter $e_{\rm{d}}$ is strictly 
constant and cannot be a function of $z$, as implemented by \citet{TB1993} and \citet{SS2000}. 
While smaller values of e$_{\rm d}$ would result in a thicker disk \citep{ST2001}, we adopt $e_{\rm{d}}$
= 0.95 in all models, in order to produce a gaseous 
disk with a finite radial extent. Hence the ensemble \emph{mean} density of warm gas is given by: 

\begin{equation}\label{warm}
\frac{\overline{\rho}_{\rm{disk}}(r,z)}{\overline{\rho}_{\rm{disk}}(0,0)} =
\exp\left[-\frac{\Phi_{\rm{tot}}(r,z) - e_{\rm{d}}^2\Phi_{\rm{tot}}(r,0) -
    (1-e_{\rm{d}}^2)\Phi_{\rm{tot}}(0,0)}{\sigma_{\rm{t}}^2 + c_{\rm{s,d}}^2}\right]
\end{equation}.

Figure \ref{cont} shows density contours of various homogeneous ISM distributions,  demonstrating
the effect of varying the parameters $\sigma_{\rm{t}}$ and $T_{\rm{d}}$. The
case of $\sigma_{\rm{t}} = 0~ \rm km ~s^{-1}$ is equivalent
to the density distribution used by, for example, \citet{SS2000}. In this case,
the average  temperature of the disk must be high 
(~$10^{5} -10^{6}$ K) in order to produce reasonable disk scale-heights
(bottom row). However, this scenario is unrealistic as gas at this
temperature rapidly cools. The turbulence parameter
$\sigma_{\rm{t}}$  allows us to increase the disk scale-height, whilst keeping
the temperature at reasonable values. Nevertheless, this parameter cannot be
made too large as the turbulence quickly becomes hypersonic, and gas in the
disk would be highly dissipative. In these simulations we set the average
temperature  of the disk gas to be $T_{\rm d} = 10^{4} ~\rm K$ and we use the
cloud velocity dispersions of  $\sigma_{\rm{t}} = 60 \rm~and~75~km~
s^{-1}$ (panels a and b respectively). 

Whilst these values are supersonic with Mach numbers of the order of 5-6, they
were chosen in order to obtain a reasonable disk thickness without the gas
being excessively supersonic and, as noted, to avoid excessively high
temperatures. The supersonic turbulence may be driven by star formation; it is
also possible that the vertical pressure support is provided by magnetic
fields. Note however, that supersonic velocities in gaseous disks are not
unknown. For example, the disks in M87 and NGC~7052 are inferred to be
supersonic with velocity dispersions $\sim 200 \> \rm km \> s^{-1}$ in M87 and
up to $400 \> \rm km \> s^{-1}$ in NGC~7052
\citep{Detal1997,VV1998}. Moreover, the concept of supersonic
turbulence is not new in the context of starburst galaxies and is a key
feature of the simulations of the ISM in such galaxies
\citep[e.g.][]{WN2001,WN2007}. In particular these papers focus on the
production of a log-normal ISM such as has been incorporated into our initial
data. Further justification for a log-normal ISM and its relation to
supersonic/superAlfvenic turbulence, appealing to the work of
\citet{NP1999} and \citet{PN1999} is provided in \citet{SB2007}.

The funnel seen in earlier simulations which utilize a
similar potential \citep[e.g.][]{TB1993} is still present in our model, most
noticeably in panels c and f. This is more apparent if one examines the
density distribution over a larger spatial range. This is an unavoidable
consequence of this type of disk model and one would need to revisit models
for the initial data in order to eliminate it in a physically acceptable
fashion. This is beyond the scope of this paper.

A specific inhomogeneous ISM (out of an ensemble of such possible ISMs) 
is obtained by multiplying equation (\ref{warm}) by a fractal distribution,
which has log-normal single point statistics and a Kolmogorov density
spectrum \citep[see][~for further details]{SB2007}. The fractal distribution
has mean $\mu$ = 1.0 and variance $\sigma^2$ = 5.0, where the variance
measures the concentration of mass within the dense clouds
\citep{FDS2003,SB2007}. The temperature of the disk clouds is determined by 
the pressure of the disk gas and the density of the clouds. The maximum temperature of the 
disk gas is set to be $T_{\rm{d}}$ = 3.0 $\times$ 10$^4$ K in order to prevent disk gas from
having temperatures near the peak of the cooling function. Gas at temperatures above this 
limit is replaced by hot halo gas.

In principle, the resulting warm gas distribution is supported in the gravitational potential 
by a combination of thermal pressure and turbulence. However, we did not impose a turbulent 
velocity field because the interaction with the wind generated by the starburst dominates in 
the vertical direction. However, we did find that a value of $e_{\rm d} = 0.95$ led to some 
radial inflow so that we compensated for this by adopting $e_{\rm d}=1.0$ (i.e azimuthal 
velocity equal to Keplerian velocity) for the velocity field only.

The rotation of the disk causes some additional (but unimportant) problems, as the boundary 
conditions used in the code are unable to handle inhomogeneous gas
rotating onto the computational grid. This results in numerical artifacts at
the boundaries, with streams of uniform dense gas appearing on the grid as the
disk rotates. These artifacts only appear at the external $x$
and $y$ boundaries and do not effect the evolution of the winds
produced in the simulations. Nor do they affect the production of the filaments.

We constrain the central density and temperature of the ISM in the disk and
halo to be in pressure equilibrium, with $P/k$ = 10$^6$~cm$^{-3}$ K,
consistent with the central regions of starburst galaxies \citep{CC1985}. The
parameters of the hot halo and warm disk are summarized in Table \ref{tab_param}.

\subsubsection{The Starburst Region}

The starburst region of M82 is believed to be a flattened disk
\citep[see][~and references therein]{SB1998}. We therefore adopt a cylindrical starburst
region of radius $r_{\rm{sb}}$ and height $h_{\rm{sb}}$. We inject
mass and energy into this region  proportional to the initial
density $\rho$ (eq. [\ref{mass_inj}] and [\ref{energy_inj}]) so that regions
of the ISM that are likely to contain stars have a higher 
injection rate of mass and energy. Hence, the mass
injection rate per unit volume ($V$) is given by:
\begin{equation}\label{mass_inj}
\frac{dM}{dtdV} = \frac{\dot{M}\!\rho}{\int\!{\rho}dV}
\end{equation}
and the energy injection rate per unit volume:
\begin{equation}\label{energy_inj}
\frac{dE}{dtdV} = \frac{\dot{E}\!\rho}{\int\!{\rho}dV}
\end{equation}
where the integral ${\int\!{\rho}{dV}}$ is over the volume of the starburst
region. All of the injected energy is in the form of internal energy of the gas.

Mass and energy are injected continuously into each cell of the starburst
region over the course of the simulation. The parameters of the starburst
region are summarized in Table \ref{tab_param}.

\begin{deluxetable}{lccccc}
\tablewidth{17.2cm}
\tablecaption{Simulation Parameters\label{tab_sims}}
\tablehead{
\colhead{Model} & \colhead{M$_{\rm{sb}}$\tablenotemark{a} ~(10$^6$
  M$_{\odot}$)} & \colhead{$\sigma_{\rm{t}}$\tablenotemark{b} ~(km s$^{-1}$)} &
\colhead{h$_{\rm{d}}$\tablenotemark{c} ~(pc)} & 
\colhead{Grid Size (cells)} &
\colhead{Spatial Range (pc$^3$)} }
\startdata
M01 & 3.44  & 60 & 110 & 512 $\times$ 512 $\times$ 512 &  1000 $\times$ 1000 $\times$ 1000 \\
M02 & 4.16  & 75 & 135 & 512 $\times$ 512 $\times$ 512 &  1000 $\times$ 1000 $\times$ 1000 \\
M03 & 3.96  & 60 & 110 & 512 $\times$ 512 $\times$ 512 &  1000 $\times$ 1000 $\times$ 1000 \\
M04 & 3.36  & 60 & 110 & 256 $\times$ 256 $\times$ 256 &  1000 $\times$ 1000 $\times$ 1000 \\
\enddata
\tablenotetext{a}{Mass of the starburst region}
\tablenotetext{b}{Velocity dispersion of the clouds}
\tablenotetext{c}{Scale-height of the disk}
\end{deluxetable}   

\subsection{The Simulations}

Three main simulations were performed, each of which was designed to test the formation of a 
wind in different ISM conditions. The parameters of these simulations are
described in Table \ref{tab_sims}. The resolution is $512 \times 512 \times 512$ cells, 
covering a spatial extent of $1 \> \rm  kpc^{3}$. This
allows us to follow the initial formation of the wind, with sufficient
resolution to investigate the origin of the H$\alpha$ and some of the X-ray
emission. A fourth simulation was performed in order to test the
effect of resolution. This simulation uses a smaller computational
grid of 256 $\times$ 256 $\times$ 256 cells, covering the same $1 \> \rm 
kpc^{3}$ spatial range. The simulations encompass both  hemispheres
of the wind, with the starburst region at
the center of the computational grid. Each simulation covers a time frame of
2 Myr, and thus we consider only initial stages of the evolution of a wind.

\begin{enumerate}
\item Model \textbf{M01} is the standard model, using the parameters given in
  Table \ref{tab_param}.
\item Model \textbf{M02} is the same as M01 except for the turbulent velocity of the
  clouds, which has been increased to $\sigma_{\rm{t}}$ = 75~km s$^{-1}$ to produce a thicker
  disk.
\item Model \textbf{M03} is the same as M01, but with a modified 
cloud distribution in the disk.
\item Model \textbf{M04} is the same as M01, but has a lower resolution.
\end{enumerate}

\begin{figure}[t]
\epsscale{1.0}
\plotone{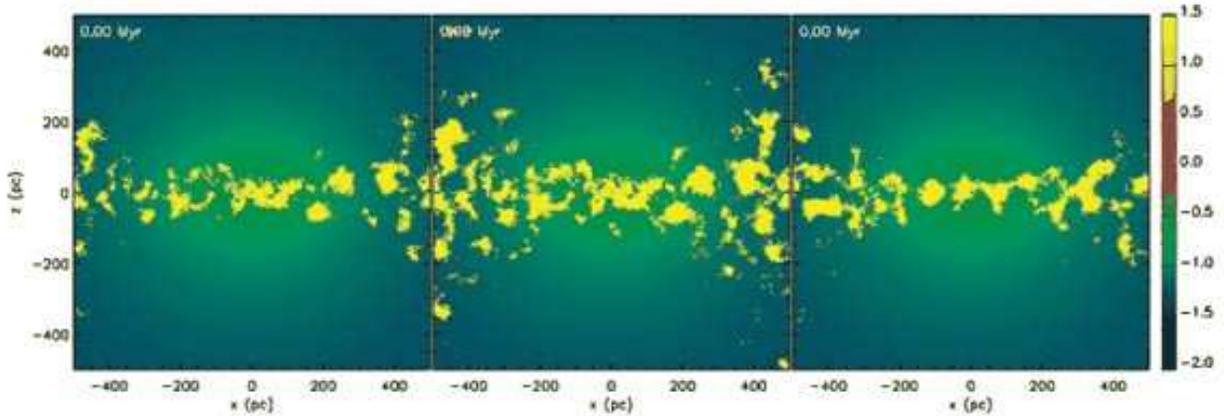}
\caption{Initial density distribution through the central y=0 plane in the three main models. (i) 
The standard model M01 (left panel), (ii) The thicker disk in M02 (center panel), and (iii) The modified
 cloud distribution in M03 (right panel).}\label{f3}
\end{figure}

The differences between the initial conditions for the three main models are illustrated 
in Figure \ref{f3}, which shows the initial density distribution through the
central y=0 plane. The model M02 was designed to produce a more collimated outflow, while M03 was
designed to test the dependence of the morphology on the inhomogeneity of the
ISM. A summary of the simulation parameters is given in Table
\ref{tab_sims}. The altered ISM distributions result in minor differences in
the mass contained within the starburst region (M$_{\rm{sb}}$) and
consequently lead to differences in the distribution of mass and
energy injected into the region.


In order to investigate the formation and structure of the filamentary
gas, we  define the H$\alpha$ emitting material in the
simulations to be gas originating from the disk and whose temperature evolves to being within the 
range of $T = 5 \times 10^3$ to  $3 \times 10^4 \> \rm K$, using a tracer variable to follow disk gas. 

\section{FORMATION OF A WIND}\label{evolution}

\subsection{Evolution, Structure and Morphology}\label{wind}

\begin{figure}[tp]
\epsscale{0.9}
\plotone{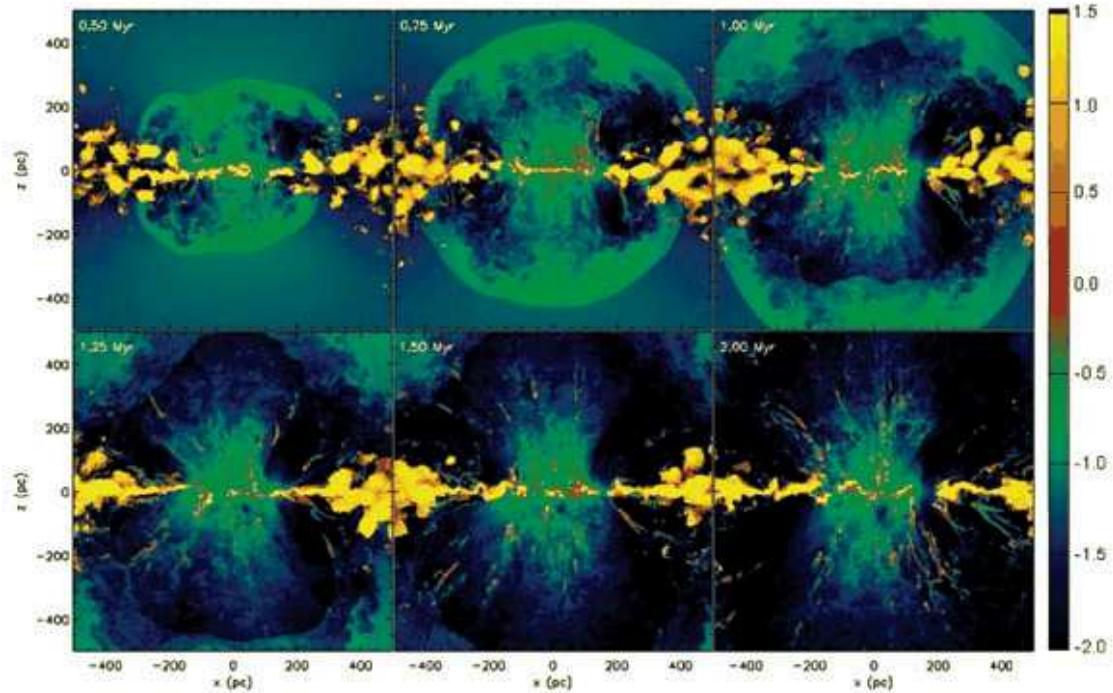}
\caption{Logarithm of the density (cm$^{-3}$) through the central y-plane of
  the wind in M01 at 0.5, 0.75, 1.0, 1.25, 1.5, and 2.0 Myr epochs.}\label{f4}
\end{figure}

\begin{figure}[t]
\epsscale{0.9}
\plotone{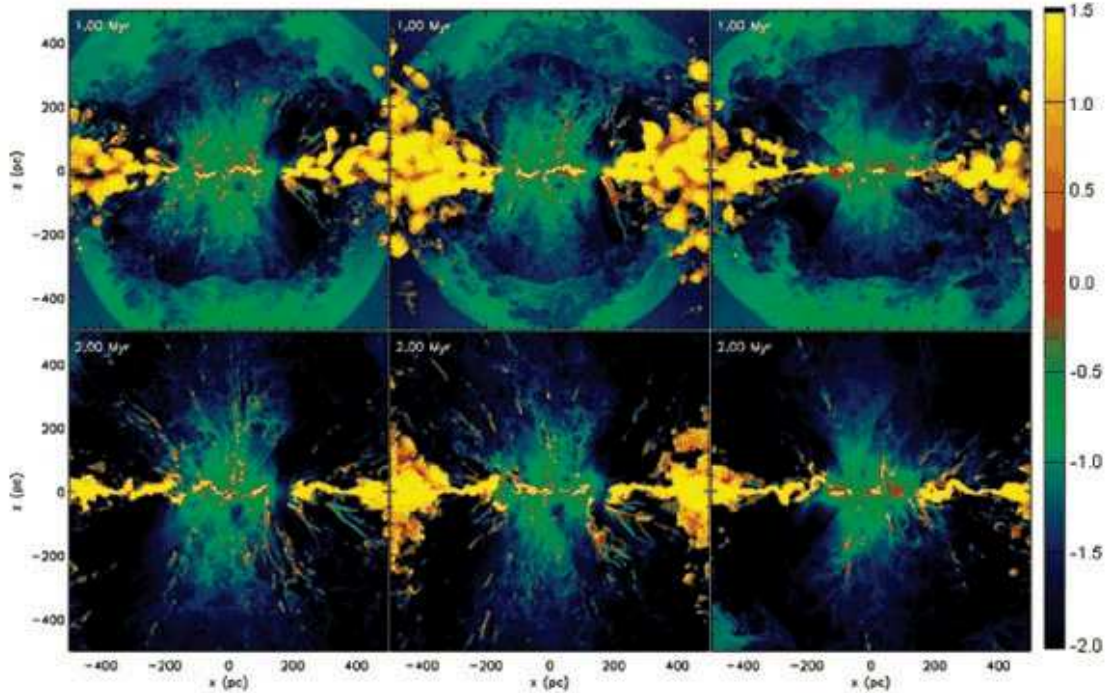}
\caption{Log-density though the central y-plane illustrating the morphology of the 
winds in M01 (left), M02 (middle) and M03 (right) at 1.0 Myr (top row) and 2.0 Myr (bottom row).}\label{f5}
\end{figure}

In this section we describe the formation and structure of the wind formed in
our main simulation M01, and discuss the effect altering the distribution
of the ISM had on the morphology of the outflow. The evolution of the wind in M01 at 6 different epochs
(0.5, 0.75, 1.0, 1.25, 1.5, and 2.0 Myr) is depicted in Figure \ref{f4}. Each frame represents the
logarithm of the density in the central y=0 plane of the computational
grid; this rendering makes much of the structure in the wind obvious. A
comparison of the morphology of the outflow at 1 Myr and 2 Myr epochs in the
models M01, M02, and M03 is shown in Figure \ref{f5}.

The wind begins as a series of small bubbles
originating from the clumpy gas in the starburst region. These bubbles merge
as they expand, forming a larger bubble that breaks out of the disk at approximately
0.15 Myr. The introduction of the inhomogeneous disk leads to a marked dependence of the
morphology of the wind on the distribution of the ISM, with the initial shape of the outflow
determined by the path the wind follows as it interacts with the dense clouds. This results 
in the asymmetrical morphology seen in all
simulations. The thicker disk in M02 acts to slow the development 
of the outflow, with the wind breaking out of the 
disk at approximately 0.2 Myr, somewhat later than the winds in M01 and M03.

It should be noted that the wind does not ``blow-out'' of the disk
as observed in numerous homogeneous simulations \citep[see for
  example,][~and references within]{MMN1989,SS2000}.
This is the result of the inhomogeneous nature of the ISM in the disk of our
model. Unlike a wind formed in an homogeneous disk, which is forced to
push its way out of the dense disk, a  wind formed in an inhomogeneous medium
follows the path of least resistance, i.e. the tenuous gas
between clouds of disk gas. The density 
of these clouds prevents them from being immediately swept-up by the outflowing hot
gas. As a result, the wind does not sweep-up
and form a dense shell of disk gas, as is found in the homogeneous case. The
formation of such a dense shell has been shown to impede the expansion of the wind
until it has reached a sufficient  distance from the
disk, where the shell begins to fragment under Rayleigh-Taylor
instabilities allowing the wind to ``blow-out''  of the shell. In the 
inhomogeneous case the wind expands freely into the more tenuous halo
gas, which is swept-up to form its  own ``shell'' around 
the outflow. This shell of swept-up halo gas is also observed in the
homogeneous case after their wind has blown-out of the disk and 
is expanding into the halo. It is unlikely that the
presence of a thicker disk in our model  would result in the formation
of a dense shell of disk gas surrounding the outflow, as any wind formed in
such a clumpy medium naturally follows the path of least resistance out of the disk.    

By 0.5 Myr (Figure \ref{f4}; upper left panel), the wind has become more
spherical as it propagates into the uniform hot halo. At this stage
the structure of the bubble consists of fast ($v$ $\gtrsim$ 1000 km s$^{-1}$),
hot ($T \gtrsim 10^{7}  \> \rm  K$), turbulent gas, surrounded by a slower (
$v \sim 300-400 \> \rm km \> s^{-1}$), cooler ($T \sim 3 \times 10^{6}\> \rm K$), 
dense shell of swept-up halo gas. Aside from the slower development of the wind in M02 
and a slight difference in overall shape, the morphology at this time is similar in all models.  


\begin{figure}[t]
\epsscale{0.8}
\plotone{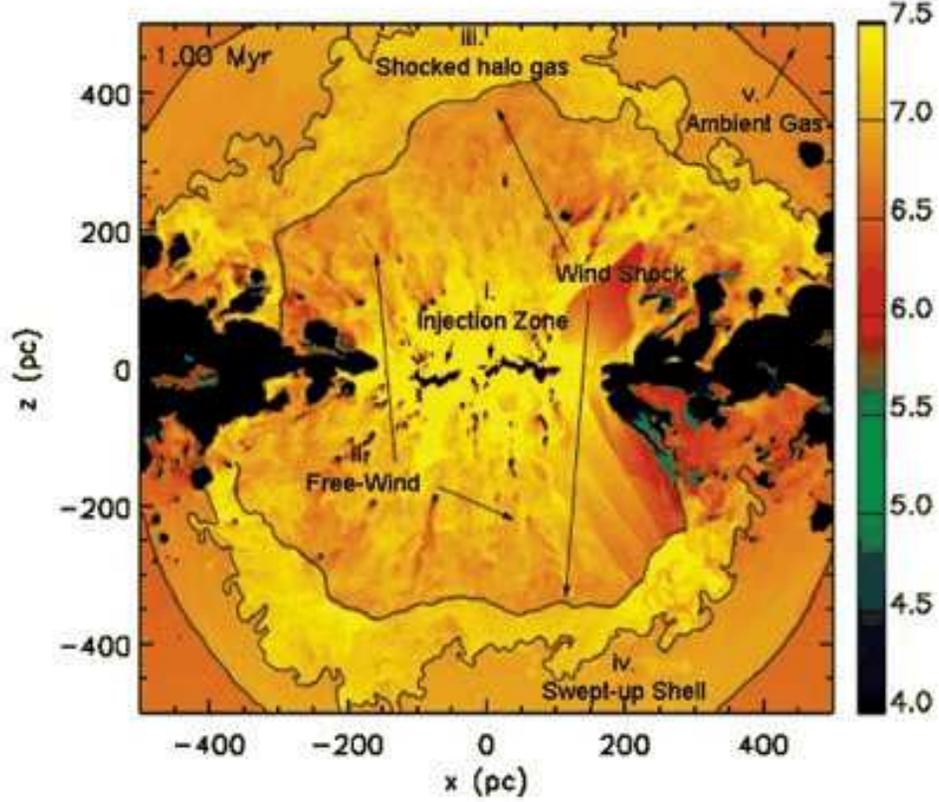}
\caption{Log-Temperature through the central y plane of M01 at 1 Myr
  indicating the 5 characteristic zones of a superwind in the ``snow-plow''
  phase of its evolution.}\label{f6}
\end{figure} 

At 1.0 Myr, the wind has begun to flow off the edge of the
computational grid. At this stage, in all simulations (Figure \ref{f5}, upper
panels) the outflow resembles the basic
structure of a superbubble in the ``snow-plow'' phase of its evolution
\citep{TI1988}, wherein the bubble is sweeping a substantial amount of the ambient hot ISM. 
This phase is characterized by 5 zones, which are illustrated
in Figure \ref{f6}:
\begin{itemize}
\item[(i)] The central injection zone (T $>$ 5 $\times$ 10$^{7}$).
\item[(ii)] A supersonic free-wind ($T \gtrsim$ 10$^{6} \> \rm  K$, $v \sim
  1000 - 2000  \> \rm km \> s^{-1}$)
\item[(iii)] A region of hot, shocked, turbulent gas (T $\gtrsim$
  10$^{7}$ K, $v$ $\sim$ 500-1200 km s$^{-1}$).
\item[(iv)] A cooler, dense outer shell  ($T \sim 7 \times 10^{6} \> \rm K$, $v \sim
  300-400 \> \rm km s^{-1}$ ).
\item[(v)] The undisturbed ambient gas ($T =  5 \times 10^{6} \> \rm K$). 
\end{itemize}
The second and third zones are separated by a wind shock. At this time, the cool, dense 
disk gas has begun to be accelerated into the wind and is distributed throughout the free-wind
region. The amount and distribution of this gas varies between models. Again,
the structure of the outflow is heavily influenced by the initial distribution
of clouds in the disk. As a result of the thicker disk, the outflow in M02 is still
somewhat less extended than the outflow in M01, but displays similar structures,
such as the ``cavity'' to the upper right of the starburst region. In M03, the
altered cloud distribution in the disk results in a somewhat different
morphology to that of M01, with the outer swept-up shell and wind shock being
wider and more flat.

At later times, most of the outer shell has flowed off the computational
grid and the shape of the wind shock has taken on a more hour-glass-like
appearance. The wind shock appears more asymmetric in the case of M03, while M02
displays similar, but less evolved structure to M01. For example, the arc of
dense disk gas to the lower right is evident in both outflows. In all models,
disk gas continues to be accelerated into the
free-wind. By 2.0 Myr (lower panels), the computational grid is mainly occupied by
the free-wind region. At this stage, disk gas that has been swept into the flow
forms filamentary-like structures, consisting of strings of clouds with
velocities in the range of $v \sim 100 - 800 \> \rm km \> s^{-1}$ (Figure
\ref{f9}).

In view of the above descriptions at various epochs, it is apparent that inhomogeneities 
in the disk result in asymmetries on relatively small ($\sim 100 \> \rm pc$) scales. However,
in all models the wind becomes more uniform as it propagates into the homogeneous
halo. It is therefore likely that asymmetries on the large scale \citep[e.g. tilted outflows;][]{BC2003} 
are caused by inhomogeneities in the halo gas, such as the neutral hydrogen cloud enveloping 
M82 \citep{YHL1993,YHL1994}.

\subsection{Base Confinement}

A difficulty with previous two-dimensional simulations of starburst winds is their inability
to confine the base of the outflow, which expands  over the course of the
simulations, resulting in unrealistic base
diameters \citep[e.g.][]{TB1993,Setal1994,SS2000}. For example, the simulations
of \citet{SS2000} resulted in base diameters of the order of 1000 pc, whereas
the base of M82's outflow has been observed to be $\sim$ 400 pc
\citep{SB1998}. While the base diameter of $\sim$ 600-800 pc for M82 observed by
\citet{Greve2004} compares more favorably to those found in previous
simulations of these winds, it is clear that some mechanism must be in place
to prevent the base of the outflow from expanding radially as the wind
evolves.

\citet{TM1997,TM1998} were able to confine the base of the outflow in their simulations by including the
inflow of disk gas onto the nucleus of the galaxy. This resulted in the downward ram pressure of the 
infalling gas to  be greater than the thermal pressure in the central region, preventing the outflow 
from expanding. However, as noted by \citet{SS2000}, the amount of gas
required in this scenario is unrealistic.
In our simulations the base of the outflow is well confined, not expanding
beyond a radius of $\sim$ 200 pc over the 2 Myr time frame (see Figure \ref{f4}). While it is
possible that the size of the base may increase
if the simulations were followed to later times, the density of the disk gas
($\sim$ 100 cm$^{-3}$) is large enough to
impede the expansion of the outflow along the major axis of the disk.   

\subsection{Wind Collimation}

In discussing the degree of collimation of the winds, we refer to the opening
angle of the cone defined by the H$\alpha$ filaments, $\theta_{\rm H\alpha}$
and the opening angle of the cone defined by the hot wind, $\theta_{\rm HW}$.
Previous simulations of these winds \citep{TB1993,Setal1994,SS2000} were
unable to collimate the outflowing gas to the degree observed in M82
\citep[$\theta_{\rm H\alpha} \approx 30^\circ$:][]{Getal1990,Metal1995,SB1998}. 
In our simulations, the thicker disk in M02 provides the greatest
degree of collimation to the outflow with $\theta_{\rm HW} \sim
100^{\circ}$, whereas in models M01 and M03 $\theta_{\rm HW}\sim
160^{\circ}$.
 
While the hot wind is poorly collimated when compared
to the degree of collimation in M82 defined by the H$\alpha$ filaments, the
morphology of the H$\alpha$ emitting material in our simulations 
compares somewhat more favorably. In M02, the filaments are collimated to the greatest
extent with $\theta_{\rm H\alpha} \sim 60 - 70^{\circ}$. The varying cloud distributions in the
disks of M01 and M03 collimate the filaments to different degrees. Both models possess the same 
disk scale height, yet the filaments in M03, which has a more sparse
distribution of clouds, are less collimated ($\theta_{\rm
  H\alpha} \sim 80 - 90^{\circ}$) than those in M01 ($\theta_{\rm H\alpha}
\sim 70 - 80^{\circ}$). However, these value are still far larger than the
degree of collimation found in M82 ($\approx 30^\circ$).

On the basis of our simulations, we conclude that the amount of gas surrounding the starburst 
region is an important contributor in determining the degree of collimation of the outflow. 
In addition, as noted earlier (see \S~\ref{wind}), it is known that in M82
the extent of the cold gas surrounding the source is much more extended than
we have modeled here with a turbulent disk \citep{YHL1993,YHL1994}. This gas will also have an
important additional effect, possibly providing the additional collimation
required in M82.

\section{FILAMENTARY H$\alpha$ EMISSION}

\subsection{Formation of the Filaments}\label{filaments}

\begin{figure}[t]
\epsscale{1.0}
\plotone{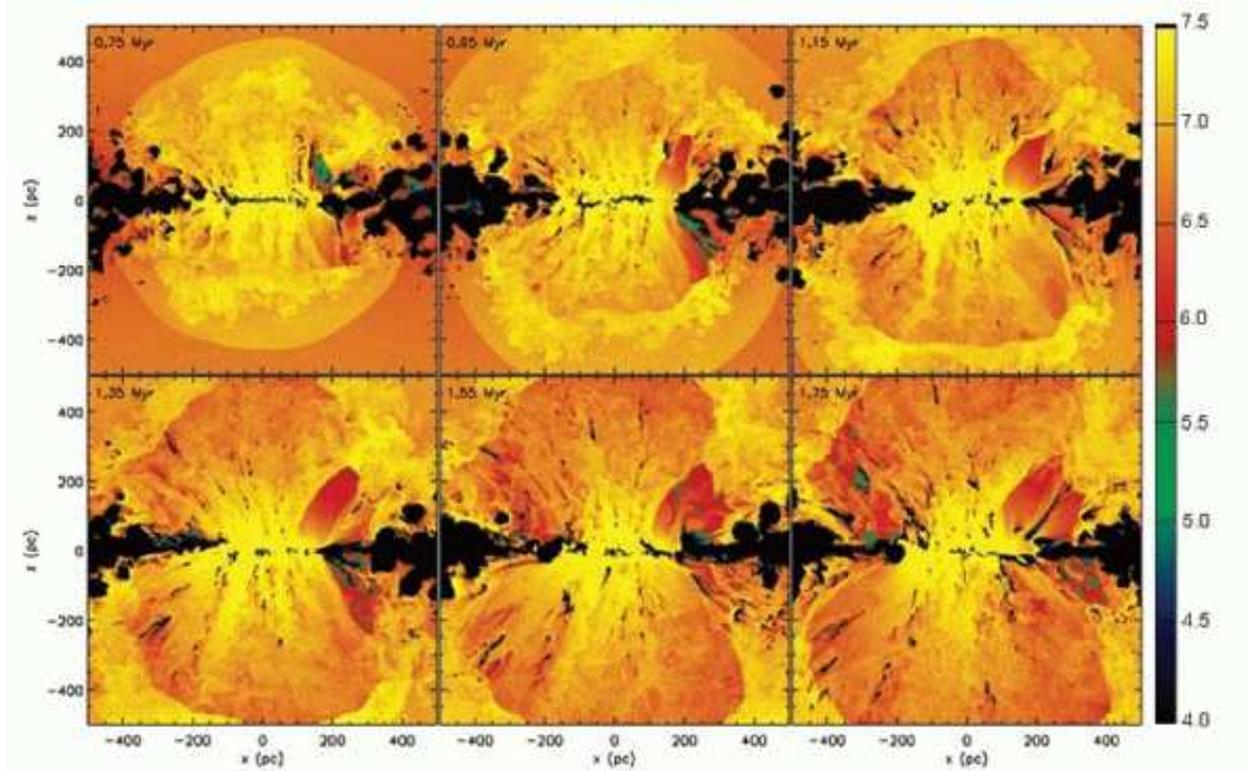}
\caption{Logarithm of the temperature (K) through the central y plane of M01 at 0.2 Myr
  intervals between 0.75 and 1.75 Myr, illustrating the formation of the filaments. }\label{f7}
\end{figure}

Emission line filaments are a dramatic feature in the images of starburst galaxies so that 
there is a large amount of interest in the mechanism behind their formation. Figure \ref{f7} shows 
log-temperature slices through the central y=0 plane of M01 over the period of
0.75 to 1.75~Myr. The filaments appear as dense clouds of disk gas that has
been drawn into the flow. Since energy is injected 
into the starburst region proportional to the local density, a significant fraction of 
the energy is injected into the dense clouds which appear in the log-normal, fractal distribution. 
The binding energy of clouds is quickly overcome and the H$\alpha$ filaments then form from the
break up of these clouds, the fragments of which are then accelerated into the
flow by the ram-pressure of the wind. 

This process is illustrated in Figure \ref{f7}, where at 0.75 Myr (upper
left panel) the starburst region  is filled with clumped
disk gas. Over the next 1 million years, the
break-up of the central clouds can be seen, with material being
drawn out into strings of dense clouds (lower panels). By 1.75 Myr, the
starburst region is almost completely evacuated. The filaments are initially
immersed within the turbulent hot gas in the vicinity of the starburst
region. As the wind expands, the filaments are accelerated into the free-wind
region of the outflow. Gas is also stripped and entrained into the wind from
clouds at the edge of the starburst region. (See in particular the panels from 1.15~Myr onwards.)

\begin{figure}[t]
\epsscale{0.8}
\plotone{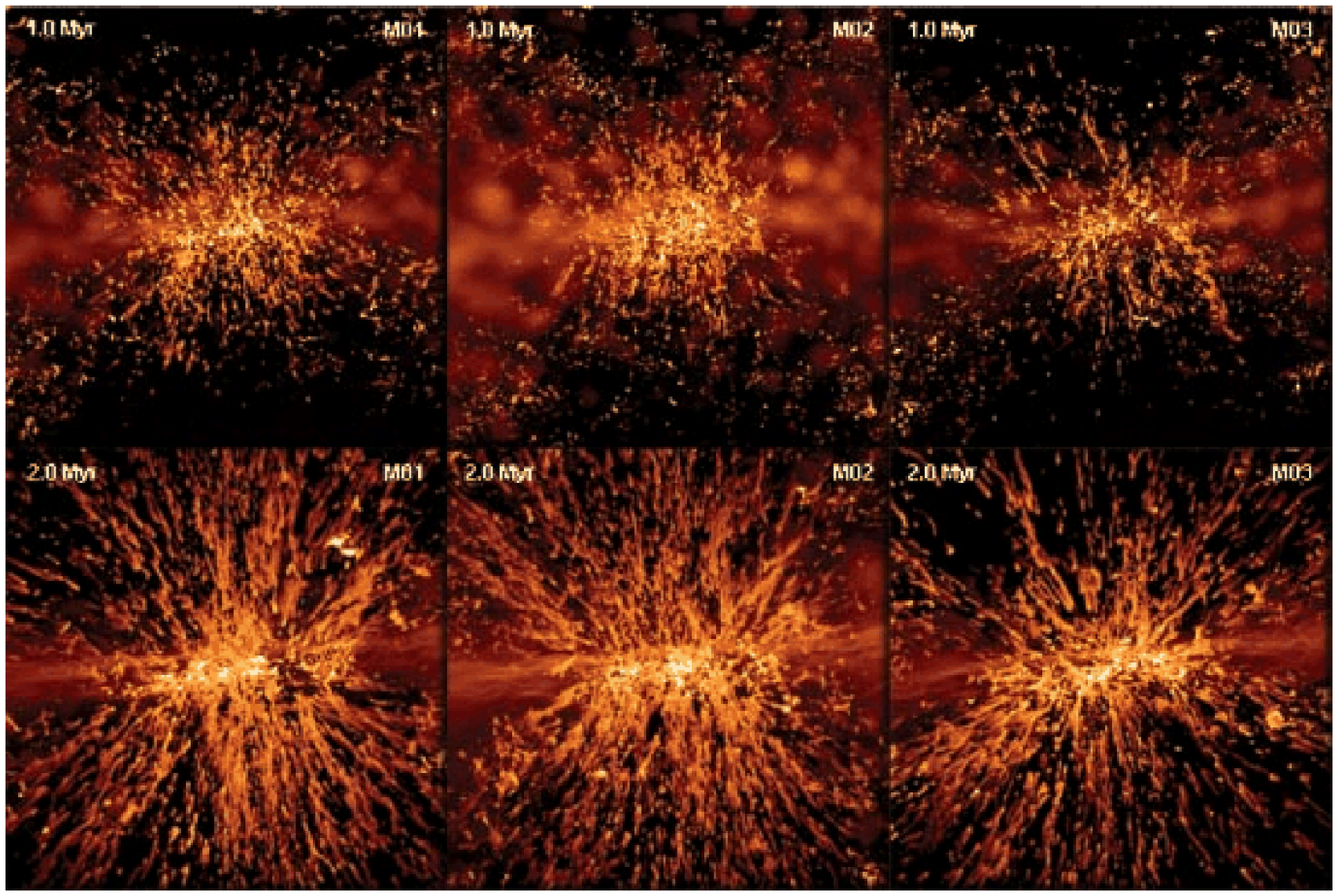}
\caption{Volume renderings, utilizing a depth cue average, of the H$\alpha$
  emitting gas in the three models at 1 Myr (top row) and 2 Myr (bottom row). } \label{f8}
\end{figure}

\begin{figure}[t]
\epsscale{1.0}
\begin{center}
\plottwo{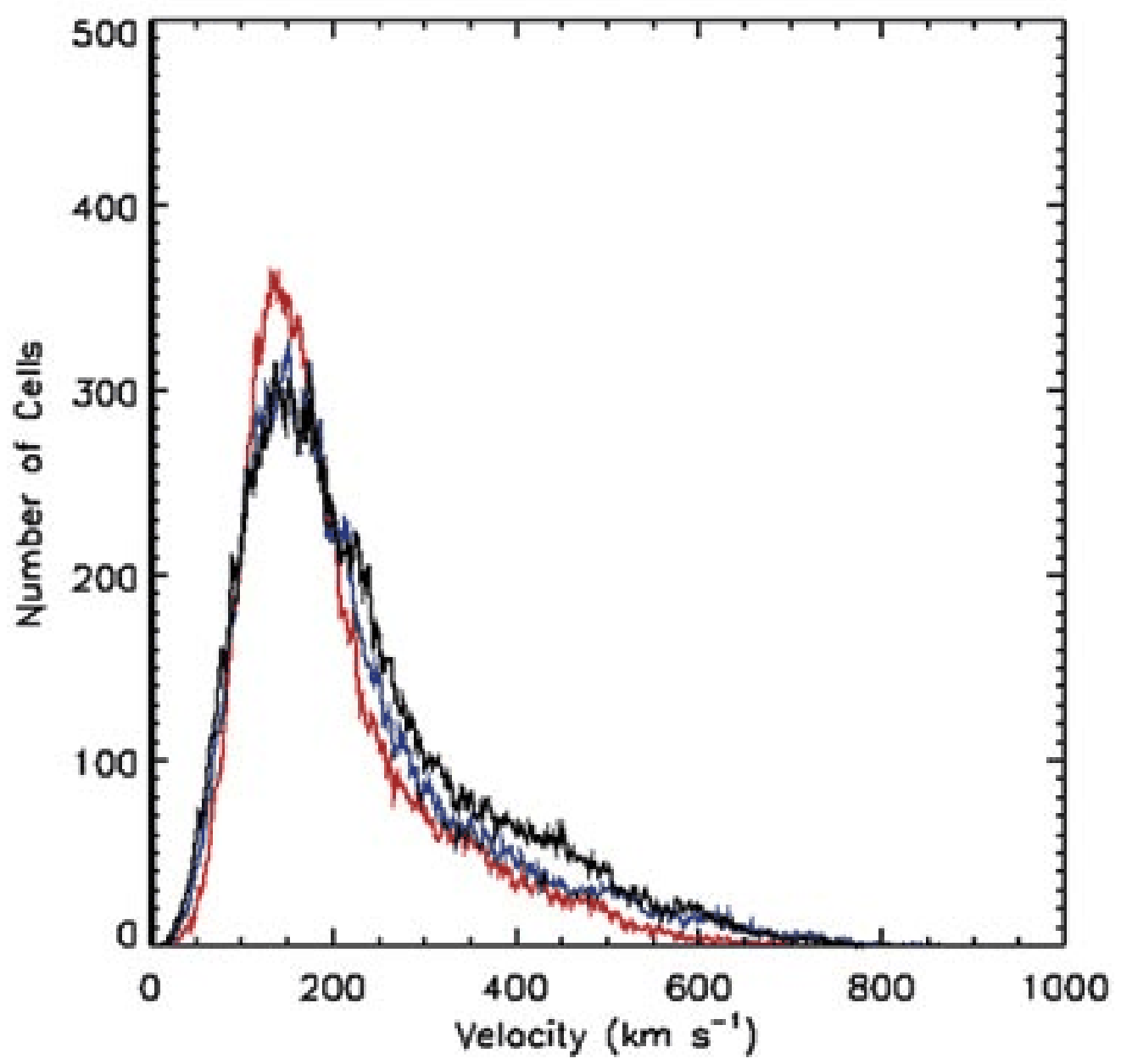}{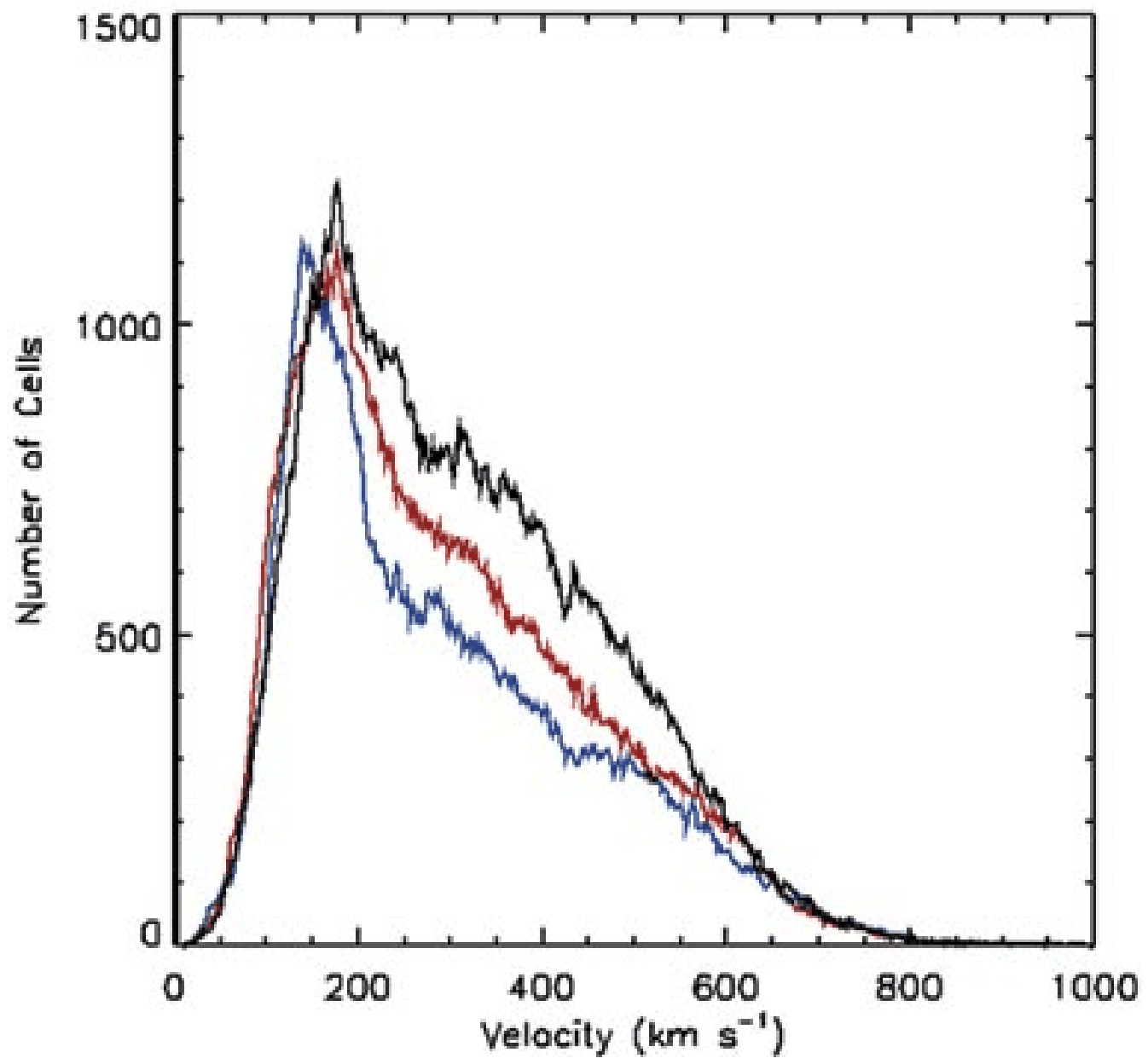}
\caption{Histogram of the velocity at 1~Myr (left) and 2~Myr (right) of the
  H$\alpha$ emitting gas in the M01, M02 and  M03 (Online: Black, red and blue
  respectively)} \label{f9}
\end{center}
\end{figure}

The filamentary structure can be seen best in Figure \ref{f8} 
which shows the three-dimensional structure of the H$\alpha$ emitting gas at 1 Myr (upper panels) 
and 2 Myr (lower panels) in M01, M02 and M03.  The filaments appear as strings of
dense clouds emanating from the starburst region. These filaments form
a filled biconical structure inside of the more spherical hot wind, with the filaments
distributed throughout this region. In the online version of this paper, the
formation of the filaments in M01 is animated in Figure \ref{fig16}. At 2 Myr
the  velocity of the H$\alpha$ emitting gas in all models
falls within the range  $v \sim$ 100-800~km $s^{-1}$
(see Figure \ref{f9}) and increases with height $z$ above the disk. This is
comparable to the  velocities observed in M82, at a height of 500 pc, of $v$ =
500-800~km $s^{-1}$ \citep{SB1998,Greve2004}. It is possible that lower
velocities are not observed  in the filaments of M82 because of dust obscuration of the central 
starburst \citep{Detal2000}.
 
\subsection{Filament Survival}

The interaction of a cloud of dense gas with a supersonic wind has been
investigated in the past via numerous two- and three-dimensional simulations
\citep[e.g.][]{Sgro1975,KMC1994,PFB2002,MD2004,MD2006,MDR2005,Metal2005,Petal2005,Tetal2006}. A
common theme in this work is the issue of the survival of the cloud which is
subject  to shock disruption. Another effect to consider is the ablation of
the cloud as  a result of the Kelvin-Helmholtz instability. This is discussed
by \citet{KMC1994} and  also further below. In adiabatic simulations
\citep{KMC1994,PFB2002} clouds  are heated and disrupted on a shock-crossing
timescale. The heating  and expansion of the cloud renders it susceptible to
ablation by the  surrounding stream. However, as explained by
\citet{MDR2005} cooling effects  a dramatic difference to the adiabatic
scenario: If the  cooling time is short enough (e.g. compared to the
cloud-shock crossing time)  then the radiative shock driven into the cloud
provides a protective  high density shell which prevents further
disruption. 

In view of the importance of cooling, we compare the cooling time to both the
cloud-crushing time  $t_{\rm crush} \approx R_c/v_{\rm sh} \approx \left(
\rho_c / \rho_w \right) R_c / v_w$  and the Kelvin-Helmholtz timescale
$t_{\rm   KH} = R_c(\rho_c+\rho_w)/(v_c-v_w)(\rho_c\rho_w)^{1/2}$,
in order to ascertain  whether we can expect the clouds to be protected by an 
enveloping radiative shock.

The cooling time of a R$_c$ = 5 pc cloud in our simulations is of the order
10$^{10}$ seconds. This is far shorter than the crushing time of the same cloud
$t_{\rm crush} \sim 10^{14}$ seconds and the growth rate of the
Kelvin-Helmholtz instability $t_{\rm KH} \sim 10^{12}$ seconds, for $v_c \sim 800 \> \rm km \>
s^{-1}$. This suggests that a cloud may be accelerated to the velocities found in this study and
remain sufficiently stable to ablation. However, we note that the
Kelvin-Helmholtz timescale could be shorter for clouds at lower velocities and
that mass ablation may occur a faster rate as a result of the heating of the
clouds outer layers by photoionization \citep[e.g.][]{Tetal2006}.

A related issue is whether the clouds can be accelerated to supersonic
velocities (based on the internal cloud sound speed).  Consider a simple model
of a spherical cloud of density $\rho_c$ and radius $R_c$ being driven by a
wind of density  $\rho_w$ and velocity $v_w$. Let the drag coefficient be
$C_D=1$. Then the theoretical acceleration of the cloud is
\begin{equation}
f_{th} = \frac {3}{8} C_D \, \left( \frac {\rho_w}{\rho_c} \right) \,
\frac {v_w^2}{R_c}
\end{equation} 
If we take one of our clouds with radius $R_c$ = 5 pc and average number density $n_c$ =
100 cm$^{-3}$ then the theoretical acceleration of the cloud, by a wind with
velocity $v_w$ = 2000 km s$^{-1}$ and average number density $n_w$ =
0.05 cm$^{-3}$, is $f_{th} = 5 \times 10^{-12} \> \rm km~s^{-2}$. The
observed acceleration of a gas cloud in our simulations is $f_{ob} \approx \> 1.3
\times 10^{-11} \> \rm km~s^{-2}$ -- close to the theoretical value. This is
physically feasible as a result of the protection by the radiative
shell. Nevertheless, this estimate neglects detailed hydrodynamics including
the formation of shocks and the ablation of material, which are important in a
realistic wind-cloud interaction. Detailed higher resolution simulations of a
single cloud impacted by a wind are necessary in order to investigate this
problem in sufficient detail. These simulations are currently in progress.

Other previous simulations have also addressed similar situations. For
example, \citet{Tetal2006} have investigated clouds
driven by an  outflow from a central star cluster and found similar ablated
cloud morphologies to those presented in this paper. In their case the clouds
only achieve a maximum velocity of $\sim 50 \> \rm km \> s^{-1}$ but the
theoretical accelerations are similar. This again points to the requirement of
high resolution simulations and the dependence on initial conditions in order
to fully understand the physics of wind-cloud interactions.

\subsection{Effect of the ISM}

As illustrated in Figure \ref{f8}, the morphology of the H$\alpha$
emitting filaments is affected by the inhomogeneity of the disk to a greater extent
than the hotter wind that surrounds them. Whilst more apparent early in
the formation of the wind, asymmetries are seen on both small and large
scales, with the distribution and number of filaments differing between the
upper and lower winds in all models.   

At 1 Myr (top row) the H$\alpha$ emission starts to become
filamentary. These filaments are asymmetric with the morphology
varying from model to model. The filaments in M02 are less extended to those
in M01. In the case of M03, the filaments to the north are tilted with
respect to the minor axis, and are overall less numerous. At 2 Myr (bottom
row) the filaments form a biconical shape,
consisting of strings of H$\alpha$ emitting clouds. These filaments are
distributed throughout the wind and rotate in the same direction as the
disk. In M02 the filaments have a similar distribution to those in M01,
but are slightly more collimated. In contrast, the filaments in M03 are more
chaotic than those in M01, with less H$\alpha$ emitting gas. Filaments are
formed from fragments of clouds in the starburst region that have
been accelerated into the wind, and the morphology of the filament system
somewhat depends on the original location of the clouds in the starburst region.

The mass of H$\alpha$ emitting gas on the computational grid in M01 and M03 at 2 Myr is M$_{\rm{H\alpha}}$ =  1.5
$\times$ 10$^6$ M$_{\odot}$ and M$_{\rm{H\alpha}}$ = 1.3 $\times$ 10$^6$
M$_{\odot}$ respectively. M02, with its thicker disk, has a H$\alpha$ mass of
M$_{\rm{H\alpha}}$ = 3.9 $\times$ 10$^6$ M$_{\odot}$ contained in the outflow. These numbers
compare favorably with known H$\alpha$ estimates in starburst winds
\citep[$\sim 10^{5}-10^{7} \> {\rm M}_{\odot}$:][]{VCB2005}. 

The amount of H$\alpha$-emitting gas found in the outflow is affected by the amount of
disk gas inside and also surrounding the starburst region. For example, the
outflow in M02 must interact with considerably more disk gas before it is able
to escape the disk. Consequently, more H$\alpha$ emitting gas is found in the
wind. However, M03 which initially contains more gas inside its starburst
region than M01 (see table \ref{tab_sims}), has a smaller amount of
filamentary gas in the wind, as a result of the more sparse distribution of clouds
surrounding the starburst region.     

\subsection{Morphology and Structure}

There have been many theories for the origin of the optical line filaments 
observed in starburst winds. A popular idea is that the
filaments are formed from disk gas that is swept up by the wind
\citep{VCB2005}. These simulations indeed confirm this idea, with the H$\alpha$ emitting gas 
forming from disk gas that has been broken up and accelerated into the
wind. However, while the filaments do form a
biconical outflow (Figure \ref{f8}), they are immersed inside the hot wind and do not trace
the true \emph{radial} and \emph{vertical} extent of the outflow as defined by
the hot gas (Figures \ref{f4} and \ref{f7}). This paints a different picture to the commonly held view of the
optical line-emission filaments framing the edges of a biconical outflow. It is possible that
current observations of starburst winds at optical and X-ray wavelengths may
not indicate the absolute size of the outflow. This has implications for
observational estimates of the energy contained in these winds, as a
significant fraction of the energy contained within the wind may be found in
the hottest (T $\gtrsim$ 10$^7$ K) gas that is not traced by H$\alpha$ and X-ray emission \citep{VCB2005}. 

Our inference of a more extensive wind than implied by the filaments is
supported by some observational evidence suggesting that the true extent of  M82's superwind is
larger than originally thought. \citet{LHW1999} find evidence for H$\alpha$ and X-ray
emission at a distance of approximately 11 kpc from the disk. They propose an
interaction of the wind with an HI cloud in the halo of
M82. This feature is now known to be connected to the main superwind emission by
X-ray emission \citep{SRB2003}, but is possibly of a different origin than the X-ray emission at lower radii
\citep{Setal2002}.  Recent Spitzer observations reveal a large mid-infrared
filamentary system along the minor axis, which is radially and vertically more
extended than the H$\alpha$ emission \citep{Eetal2006}. While the nature of
this emission is uncertain, there does appear to be a spatial
correlation with the H$\alpha$ emitting gas in the region where H$\alpha$
emission is detected. \citet{SetalA2004} also possibly detect diffuse,
low-surface brightness, X-ray emission  in in M82's halo, which has a larger
spatial extent and uniformity than the filamentary X-ray emission. On the
other hand, they note that this may be caused by low photon statistics. It is clear that multi-wavelength 
observations are needed in order to understand the true extent of starburst winds. 

The tendency of the filaments to fill the interior of the biconical structure
arising from our simulations is also of interest. This is in agreement with
observations of the wind in the Circinus galaxy \citep{VB1997}. Other winds,
such as M82 \citep{SB1998} and NGC 3079 \citep{Vetal1994}, are
limb-brightened, with the filaments thought to lie on the surface of a mostly hollow structure. 
The mechanism responsible for producing an evacuated cavity is uncertain, but in view of the way in 
which filaments have been formed in our simulations, this feature may reflect 
the distribution of the interstellar medium in the starburst region
itself. Starbursts where much of the molecular gas is  situated in a ring
\citep[e.g.][]{TDW1993}, rather than  throughout a disk, would most likely produce winds that 
are hollow, as the clouds in the ring are broken up and entrained into the
wind. Another possibility is that the wind has significantly evolved to point
where it has evacuated the center of the starburst region of molecular gas,
with the filamentary system being fed from gas stripped from the edge of the
starburst region, an effect that in fact does occur in our simulations
(e.g. Figure \ref{f7}; lower left panel). The center of the biconical region
could also have been swept clear by a previous wind, powered by an earlier
burst of star formation \citep[e.g.][]{BC2003,Fetal2003}.   
  
The source of the ionization of the filaments in a starburst wind is still
uncertain. Since photoionization is not included in our simulations, all of
the H$\alpha$ emission arises from shocks. Indeed there are examples of winds
where the emission is shock ionized \citep[e.g. NGC 1482;][]{VR2002}. In other
winds, such as M82 and NGC 253, there are signs that some of the filaments may
be photoionized. In particular, M82 is known to have a strong ionization cone,
where emission in the lower filaments is thought to arise from
photoionization, with ionization from shocks dominating at larger radii
\citep{SB1998}. While we are unable to study photoionization with our current
model, it is likely that it plays a role in the ionization of the filaments in
many winds \citep[see, for example,][]{MDR2005,Tetal2006}, and 
warrants further investigation.

\section{X-RAY EMISSION}

\begin{figure}[t]
\epsscale{1.0}
\plottwo{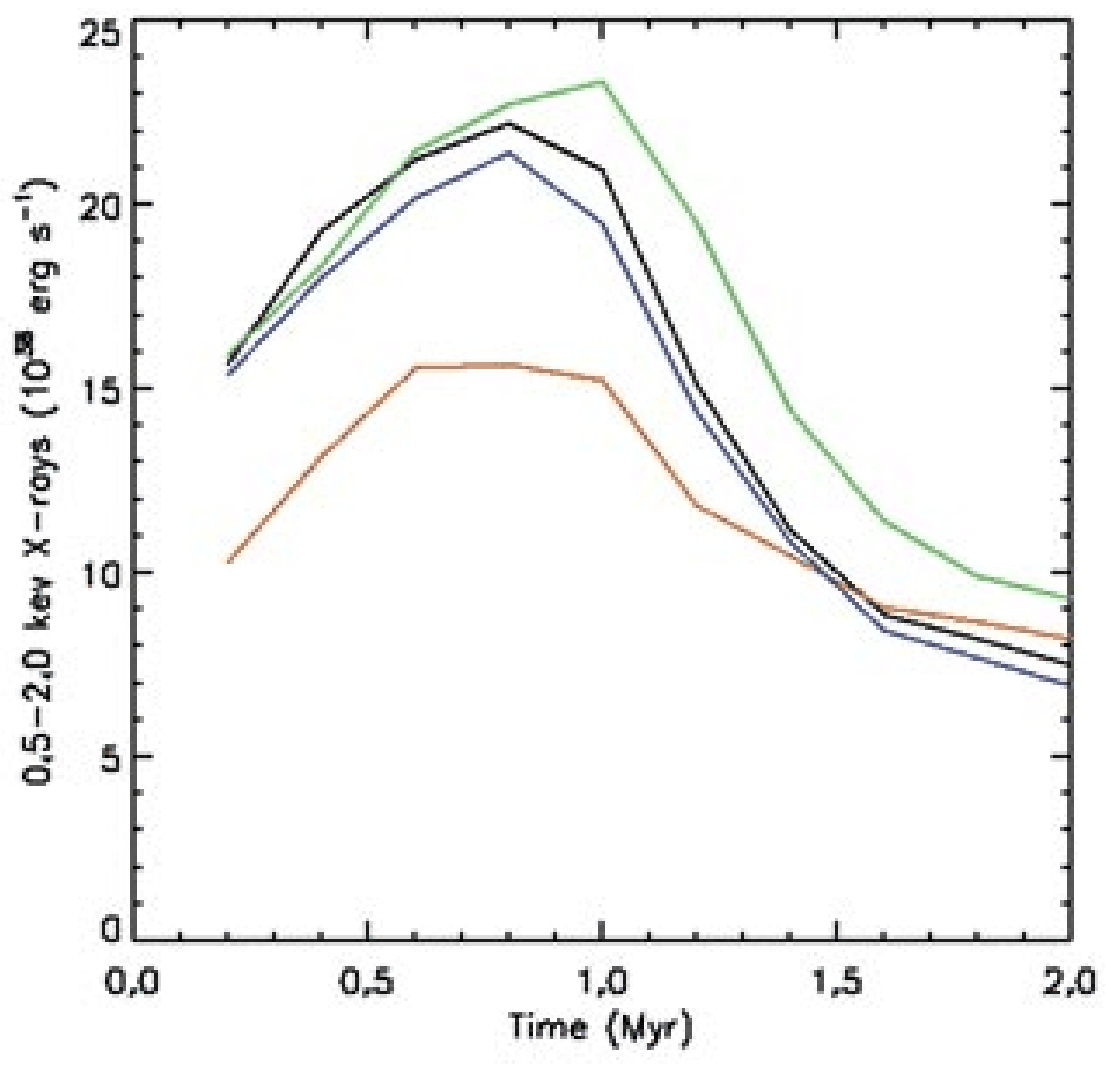}{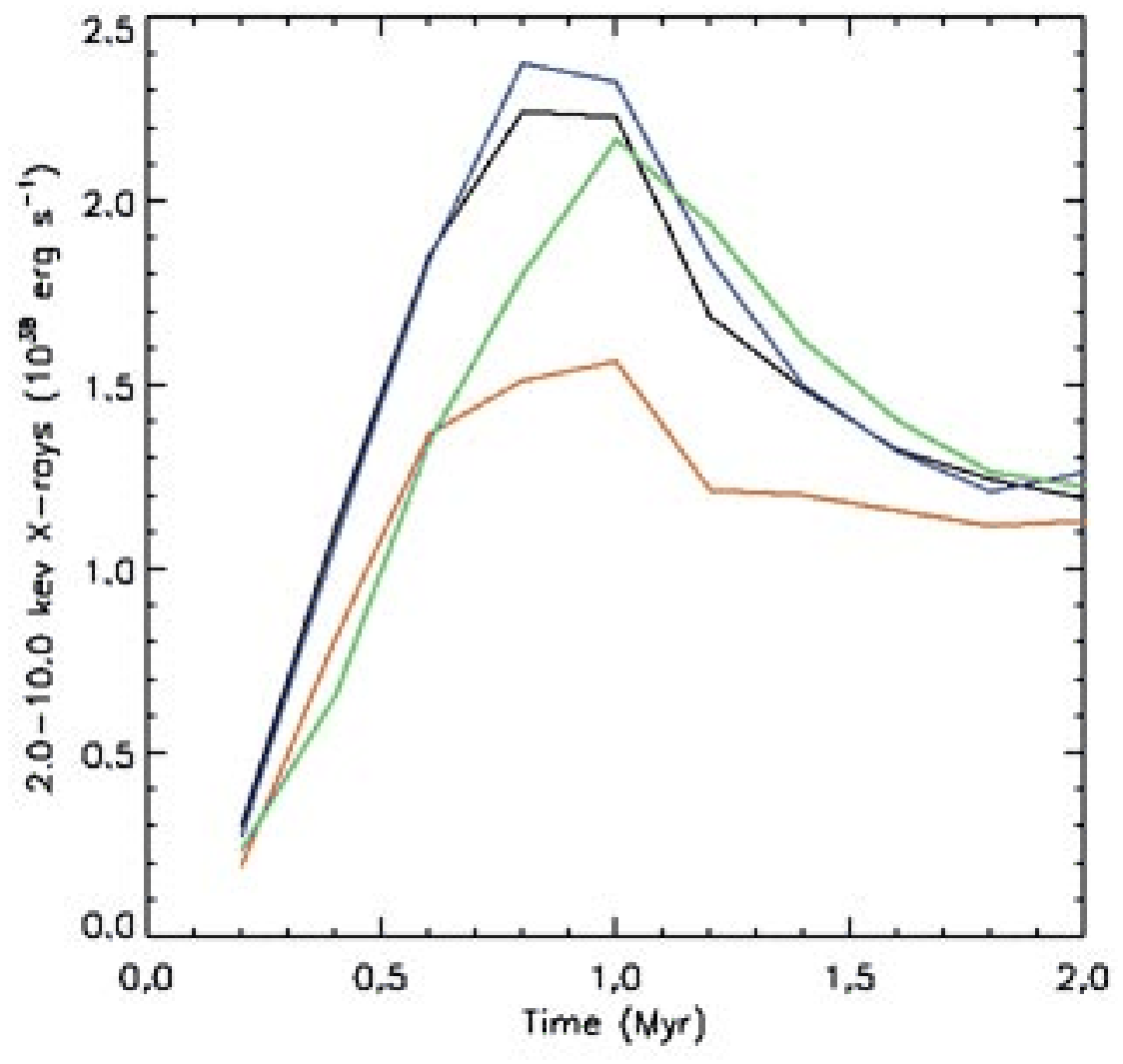}
\caption{X-ray luminosities in both soft (left) and hard (right) energy
  bands for all models (Online: M01 (black), M02 (red), M03 (blue), and M04 (green)).} \label{fig10}
\end{figure}

X-ray luminosities implied by the simulations were determined at 0.2 Myr
intervals in both the soft (0.5 - 2.0 kev) and hard (2.0 -
10.0 kev) energy bands, utilizing broadband cooling fractions obtained from
\textsc{Mappings} IIIr \citep[see~][]{SD1993}. Figure \ref{fig10} gives the X-ray luminosity of the
wind as a function of time in both energy bands for all models. The peak of the curves in
Figure \ref{fig10} is the result of the  limited 1 kpc$^3$ spatial range of the
simulations. Once the swept-up shell has reached the edge of the computational
grid (at $\sim$ 0.8 Myr) it begins to flow off the grid, and its
contribution to the X-ray luminosity can no longer be determined. 
As this happens, the curves in Figure \ref{fig10} begin to decline, flattening
when the swept-up shell has completely left the computational grid. The calculated
X-ray luminosities of the wind are then comprised solely of emission from the
free-wind region, which has grown in size to fill the computational grid.

\subsection{Hard X-ray Emission}

\begin{figure}[t]
\epsscale{0.8}
\plotone{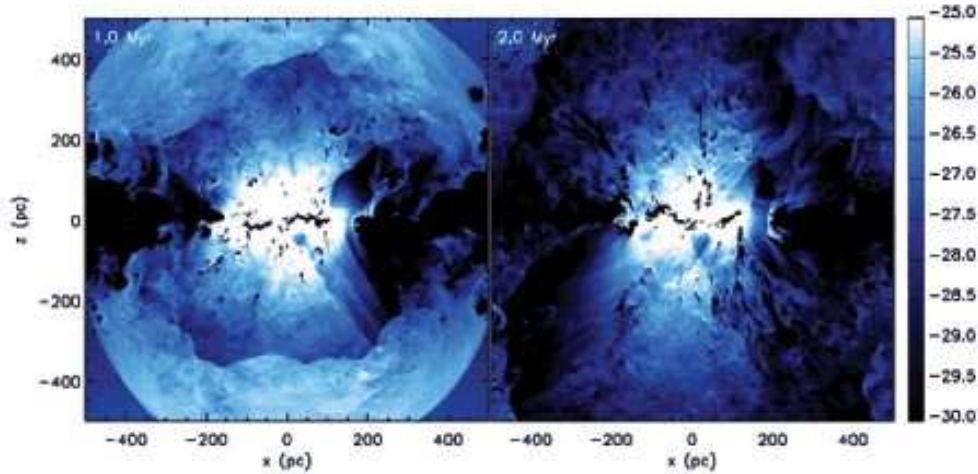}
\caption{Logarithm of the hard X-ray emissivity (erg s$^{-1}$ cm$^{-3}$) through the central y-plane
  of M01 at 1 Myr (left) and 2 Myr (right).} \label{fig11}
\end{figure}

The hard (2.0 - 10.0 kev) X-ray luminosity of the wind for all models in given in the right hand
panel of Figure \ref{fig10}. The luminosity does not vary significantly
between  models, being of the order of 10$^{38}$ erg s$^{-1}$. The hard X-ray
emissivity, through the central y-plane, of M01 at 1.0 Myr (left) and 2.0 Myr
(right) epochs is shown in Figure \ref{fig11}. At 1 Myr the
wind has started to flow of the computational grid, but the internal structure
of the wind, as shown in Figure \ref{f6}, can still be seen. The main contributor to the hard
X-ray emission is the starburst region itself, with a lesser contribution from
the swept-up shell. While the shell is not a strong X-ray emitter at hard
energies, having a temperature of the order of 10$^{6}$ K, the volume of the computational grid
occupied by the swept-up shell is large, making its contribution to the hard X-ray emission 
non-negligible, as evident by the drop in luminosity as the shell leaves the grid. Differences in the 
shape and volume of this shell in each model leads to the
variation in the peaks in Figure \ref{fig10}, with the thinner shell in M02
resulting in smaller luminosities.  

By 2 Myr (Figure \ref{fig11}, right panel), the swept-up shell has completely
left the computational grid, and the calculated X-ray luminosity is now comprised
solely from emission processes interior to the shell. The starburst
region is still the major contributor to the emission, with a lesser contribution
from emission from the more diffuse wind. This is in agreement with the
conclusions of both \citet{Setal1994} and \citet{SS2000}. Furthermore,
\citet{STA2005} developed an analytic model for the X-ray emission from star
cluster winds, which showed that the hard X-ray emission is associated with the
hot thermal plasma within the starburst region. As expected, we find
no significant difference between the hard X-ray luminosity in each
model at the 2 Myr epoch, as the size and power of the starburst is identical
in all models. At this time, the luminosity is approximately constant with L$_{\rm x}$ $\sim$
1.2 $\times$ 10$^{38}$ erg s$^{-1}$, but as the contribution from the swept-up
shell at this time cannot be determined, it is likely that the actual hard
X-ray luminosity of each model is higher, increasing as the volume occupied by
the wind increases.    

\subsection{Soft X-ray Emission}

\subsubsection{Origin of the Soft X-rays}

Chandra observations of starburst galaxies have provided clues to
the nature of the soft X-ray emission seen in galactic winds, such as a close
spatial correlation between the soft X-ray and H$\alpha$ emission in the wind
\citep[e.g.][]{SetalA2004}, which is suggestive of a physical relationship between 
filaments and the production of soft X-rays. However, the actual mechanism for the
emission of the soft X-rays is uncertain. Our simulations indicate possible
mechanisms for the  production of soft X-rays. Two of these mechanisms involve
the mixing of high  temperature gas from the hot wind and warm gas from the
filaments to produce  intermediate temperature gas which emits soft
X-rays. Whilst the resolution  of these simulations is adequate for the global
features of the  simulation, it is insufficient to resolve the fine scale
interactions between  the hot and warm gas. Therefore the production of soft
X-rays by mixing  can only be regarded as a possible mechanism at this stage.  

\begin{figure}[t]
\epsscale{0.8}
\plotone{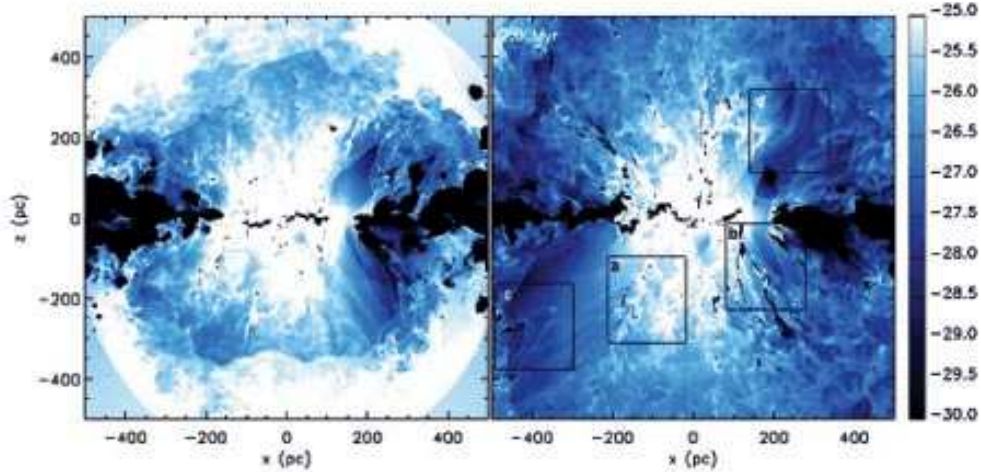}
\caption{Logarithm of the soft X-ray emissivity (erg s$^{-1}$ cm$^{-3}$)
  through the central y-plane
  of M01 at 1 Myr (left) and 2 Myr (right). The boxes indicate the location of
  regions of X-ray emission that are highlighted in figure \ref{fig13}.} \label{fig12}
\end{figure}

Notwithstanding the above qualification we discuss the regions of soft X-ray
emission that are produced  in these simulations, bearing in mind that two of
the relevant processes  described below need to be confirmed by planned
detailed simulations of interactions between winds and filaments. Figure \ref{fig12}
shows the volume emissivity of the soft X-rays at 1 Myr (left) and 2 Myr
(right) epochs. As with the hard X-ray emission in Figure \ref{fig11}, there
is little difference between the models M01, M02, and M03, with X-rays arising from the same
processes in each model. 

At 1 Myr it can be clearly seen that the swept-up shell is a strong emitter of
soft X-rays. This shell consists of halo gas that has been swept into the wind
and shock heated to  $T \sim 7 \times 10^{6} \> \rm  K$. This source of soft
X-ray emission is straightforward and is not subject to the
qualifications described above.

\begin{figure}[t]
\epsscale{0.6}
\plotone{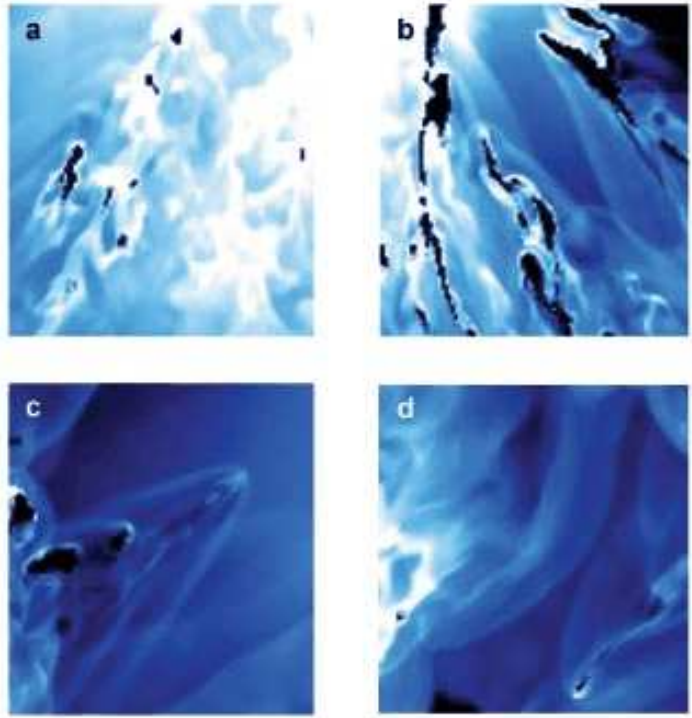}
\caption{Highlighted soft X-ray emissivity from the wind in M01. Soft X-rays
  arise from: (a) The cooling mass-loaded wind, (b) The intermediate
  temperature interface between hot and cold gas, (c) Bow shocks, and (d) The
  interaction between bow shocks. The size and location of each panel are
  indicated in Figure \ref{fig12}.}\label{fig13}
\end{figure}

At 2 Myr, X-ray emission from the free wind starts to be apparent. There are 4
main processes interior to the swept-up shell that give rise to soft X-rays and
examples of these are shown in Figure
\ref{fig13}:\\
\begin{itemize}
  \item[(a)] The mass-loaded wind. Turbulent gas in the vicinity of the
  starburst region is mass-loaded through mixing with clouds in the disk,
  creating a region of hot (T $\gtrsim$  10$^{6}$ K), dense
  (n $\sim$ 0.3 cm$^{-3}$) rapidly cooling gas. This component is a strong
  X-ray emitter and is the  largest contributor to the soft X-ray emission
  interior to the swept-up shell in all our models. As we have indicated above, numerical
  diffusion as a result of  inadequate resolution may lead to poor estimates
  of the amount of mixing  involved and consequently in the soft X-ray
  emissivity of the mixed gas.  Therefore, higher resolution simulations
  dedicated to a study of the  mixing between the hot wind and the cooler
  filaments are required in  order to correctly determine the amount of mixing involved.
  \item[(b)] Emission from the intermediate temperature ($T \sim 10^{6}$ K) interface
  between the hot wind and the  cooler filaments. This component is related to
  the mass-loaded wind,  with mixing between the hot and cold gas creating a
  region of intermediate  densities and temperatures. Again inadequate
  resolution and numerical diffusion  in this region means that the actual
  contribution to the soft X-ray  luminosity is uncertain.
  \item[(c)] Bow shocks. Soft X-rays arise when a bow shock (T
  $\sim$ 10$^{7}$ K) forms upstream of clouds of disk gas (T $\sim$ 10$^{4}$
  K) that have been accelerated into the flow by the ram pressure of the hot wind
  (T $\sim$ 10$^{6}$ K). This is a straight forward process where gas is shock
  heated to X-ray temperatures.
  \item[(d)] Colliding bow shocks. As disk gas is accelerated
  into the wind, the resultant bow shocks begin to cool as the wind
  expands. When 2 shocks collide, the gas is further shock heated to
  temperatures of the order $T \sim 10^{7}$ K.   
\end{itemize}

\begin{figure}[t]
\epsscale{0.75}
\plotone{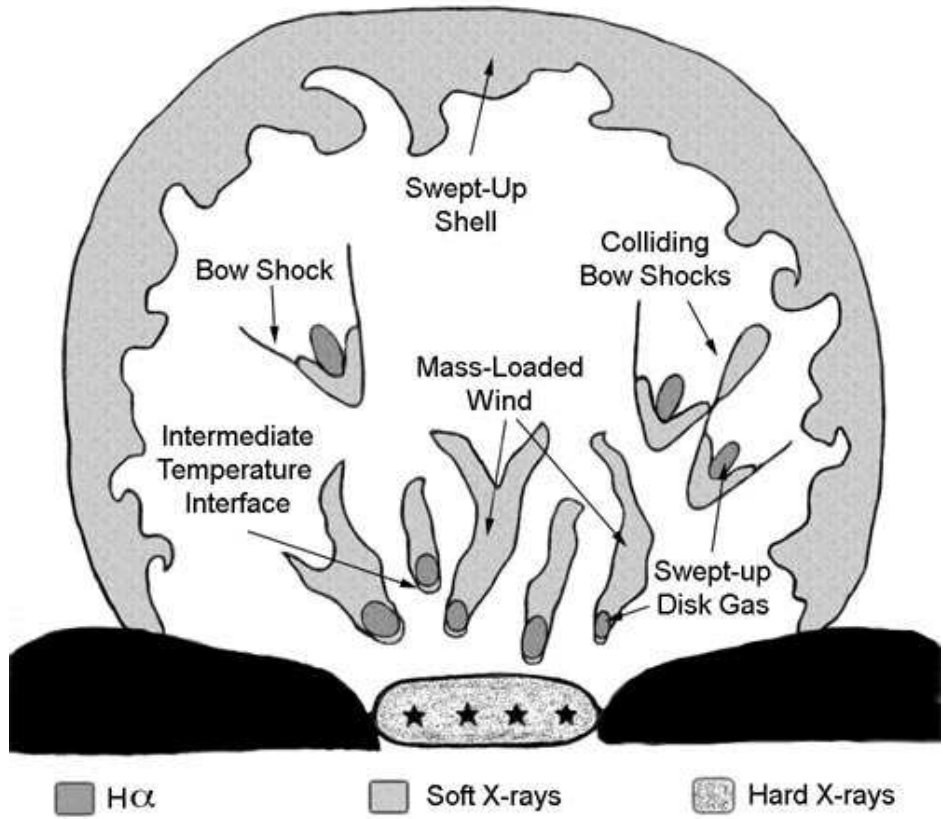}
\caption{Schematic of the H$\alpha$ and X-ray emission arising in a starburst
  wind and their spatial relationship.} \label{fig14}
\end{figure}

As a result of the limited spatial extent of these simulations, it is difficult to get
a clear picture of the distribution of soft X-ray emission throughout the
entire wind. However, it is possible to extrapolate from the structure of the
wind at 1 Myr and 2 Myr to get an idea of the soft X-ray emission arising from the wind at later times. 
Figure \ref{fig14} shows a schematic of the X-ray and
H$\alpha$ emitting gas in a starburst wind, based on the results of our
simulations. In \S~\ref{filaments} we proposed the formation of the
H$\alpha$ emitting filaments from the breakup of disk clouds in the
starburst region that are then accelerated into the wind by ram pressure. These
clouds are potentially the source of the mass-loaded gas discussed above, with tails of
soft X-ray gas streaming from their surfaces (subject to the caveats already
noted concerning numerical diffusion).  The presence of clouds of disk gas in the
outflow also results in the formation of bow shocks as they are accelerated by
the wind. X-rays that arise from these
processes are naturally spatially correlated to the H$\alpha$ emission in the
wind. While the X-ray emission in our simulations is volume
filled, which was also found to be the case with the H$\alpha$ emission (see Figure
\ref{f8}), many starburst winds are found to be limb-brightened in
X-rays. However, there is evidence to suggest that some winds may at
least be partially volume filled \citep[e.g. NGC 3079:][]{CBV2002}. Nevertheless, the
observed physical connection between the two wavelengths suggests that the same
mechanism responsible for the production of limb-brightened outflows in
H$\alpha$, result in limb-brightened X-ray emission.

\subsubsection{Soft X-ray luminosity}

We now return to discuss further features of the soft X-ray luminosity (Figure
\ref{fig10}). While the soft X-ray luminosity is similar in all of our models, slight
differences reflect the morphology of each wind. The models M01 and M03, whose
initial ISMs differed only by the distribution of clouds in the disk, have
almost identical soft X-ray luminosities. On the other hand, the wind formed in the
model M02, initially has a lower luminosity than the other models. This is a
consequence of the thinner shell that forms around the outflow. At 2 Myr, when 
the contribution of this shell is no longer taken into account, the luminosity of 
the wind M02 is comparable to that of the other models. 

The highest soft X-ray luminosities reached in our models occur when the
swept-up shell still lies on the computational grid. At this time, the
luminosity of the wind is of the order of L$_{\rm x}$ $\sim$ 10$^{39}$ erg
s$^{-1}$. When the shell is not included, and soft X-ray emission arises solely
from the processes associated with the H$\alpha$ emission discussed above, the
luminosity is of the order L$_{\rm x}$ $\sim$ 10$^{38}$ erg s$^{-1}$. Typical
soft X-ray luminosities that are observed in starburst winds fall in the
range of 10$^{38}$ - 10$^{41}$ erg s$^{-1}$
\citep[e.g.][]{RPS1997,SetalA2004,OWB2005b}. Our values fall at the lower end
of this range, but clearly the X-ray luminosity is dependent on the volume of
the wind and would be higher at later times.          

\section{RESOLUTION EFFECTS}

\begin{figure}[t]
\epsscale{0.8}
\plotone{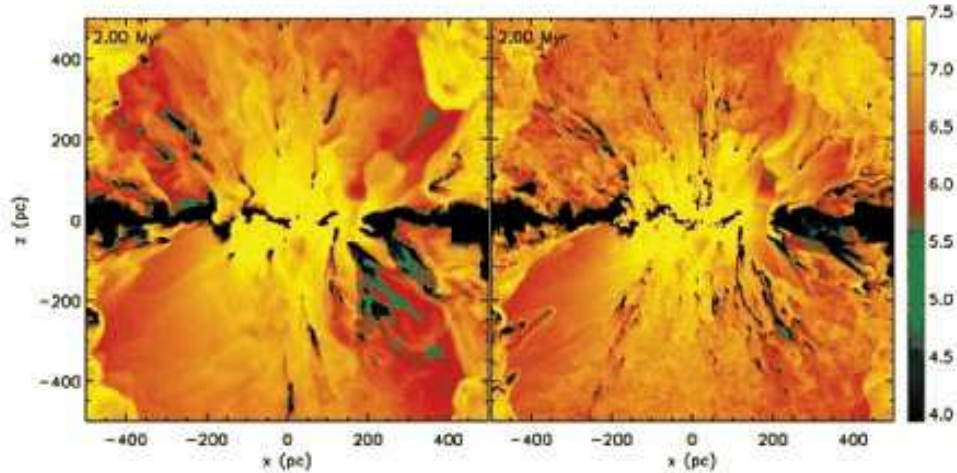}
\caption{Logarithm of the temperature (K) in the central y plane at 2 Myr in
  M04 (left) and M01 (right).} \label{fig15}
\end{figure}

\begin{figure}[p]
\epsscale{0.6}
\plotone{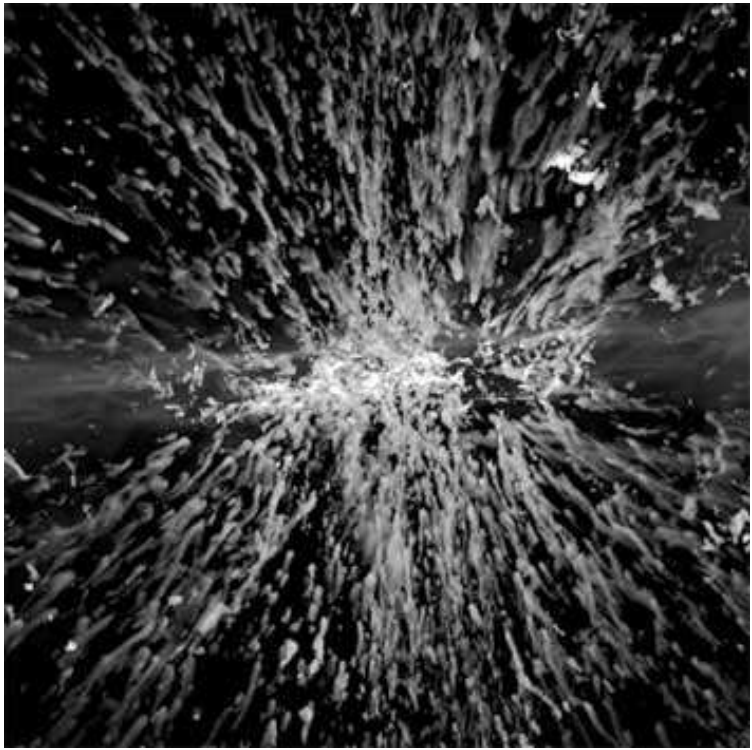}
\caption{Online: Movie the formation of the H$\alpha$ filaments in M01 over a 2 Myr time frame.}\label{fig16}
\end{figure}

In order to test the effect of the numerical resolution on our model, a fourth
simulation (M04) was performed on a smaller computational grid of 256
$\times$ 256 $\times$ 256 cells, but otherwise identical to M01. The effect of the resolution is most
significant with respect to the  H$\alpha$ emitting gas found in the
outflow. Figure \ref{fig15} gives the logarithm of the temperature at 2 Myr in M04 (left) and
M01 (right). M04 differs only by its smaller computational grid. Whilst it
exhibits a similar overall shape and structure (i.e. biconical structure),
smaller structures, such as bow shocks, are not well resolved. The dense gas
that makes up the H$\alpha$ filaments has not been adequately resolved,
appearing as strands of gas, rather than as the strings of clouds found in the
higher resolution simulations.    

We find that the lower resolution of M04 results in soft X-ray luminosities
that are comparable to those in M01 (Figure \ref{fig10}). The luminosity in
both soft and hard energy bands follows a similar trend to that of the three main
models, with the luminosity being slightly higher in the soft band and
lower in the hard. As with in the other models, the hard X-ray emission originates from the starburst
region. On the other hand, as bow shocks are not well resolved at
the lower resolution in M04, the majority of the soft X-ray emission arises
from the swept-up shell of halo gas, and from the mass-loaded component of
the wind.  

As previously noted, numerical diffusion causes
difficulties in accurately determining the soft X-ray luminosity of the
wind in regions of mixing gas. As some of the X-ray emission is poorly resolved (i.e. only a few cells
in size) it is likely that the calculated luminosities may be poorly
estimated in these regions. This is particularly likely at the interfaces
between the hot wind and the cool entrained gas, which emits strongly in soft
X-rays. This region is at best a few cells in width and any X-ray emission is
likely a result of mass diffusion between the hot and cold gas and not physical. Numerical diffusion 
may also be an issue in the mass-loaded component of the wind, which appears to be the largest 
contributor to the soft X-ray emission. Mixing of the cool gas stripped of clouds accelerated into 
the wind with the surrounding hot gas may have resulted in temperatures and densities that emit strongly 
at X-ray energies. However, we note that this effect may be physical in
origin. The fact that the soft X-ray luminosity in both M01 and M04 are similar at 2 Myr, 
when the majority of the emission is from the mass-loaded component, is encouraging, but higher 
resolution simulations are needed in order to determine if the luminosities achieved in our 
simulations are realistic.       

\section{DISCUSSION AND CONCLUSIONS} 

We have performed a series of three-dimensional simulations of a
starburst-driven galactic wind designed to test the evolution of the wind in
different ISM conditions. By conducting three-dimensional simulations we are
able, for the  first time, to study the morphology and dynamics of the entire
outflow. The introduction of an inhomogeneous disk enables us to study the
development of asymmetries and the interaction of the wind with clouds in the disk.
The results of these simulations are as follows-

\begin{enumerate}

\item The interstellar medium plays a pivotal role in the evolution of a
  galactic wind. The interaction of the wind with clouds in the disk results in
  asymmetries and tilted outflows on the small scale. Nevertheless, it is likely that
  inhomogeneities in the halo are the cause of the large-scale asymmetries in
  an outflow. 

\item The distribution of gas surrounding the starburst region assists in collimating
  the outflow. The thickness of the disk and the location of the starburst are
  important factors in determining the degree of collimation, with the degree of collimation 
  increasing with the amount of gas surrounding the starburst region. 

\item The base of the outflow is well confined within a radius of 200 pc over
  the 2 Myr time frame of the simulation as a result of the high density of the disk gas.
  
\item The H$\alpha$ filaments form from the breakup of clouds in the starburst
  region, the fragments of which are then accelerated by the ram pressure of
  the wind. Filaments are also formed from gas that has been stripped from the sides of the starburst region. 
  The distribution and mass of the filaments is affected by the distribution of
  clouds in the vicinity of the starburst region.

\item The H$\alpha$ filaments appear as strings of disk gas that form a filled
  biconical structure inside of a more spherical hot wind. The filaments are
  distributed throughout this structure, but do not trace the true extent to
  the wind defined by the hot gas.

\item The calculated soft X-ray luminosities up until 2.0 Myr are of the order of 10$^{38}$ -
  10$^{39}$ erg s$^{-1}$ and the hard X-ray luminosities of
  10$^{38}$ erg s$^{-1}$. These luminosities are dependent on the
  volume of the wind and would be larger for a more evolved outflow.

\item Interior to the swept-up shell of halo gas, soft X-ray emission
  originates in the same region as the H$\alpha$
  emitting gas. While higher resolution simulations are needed to confirm
  X-ray emission from mixing processes, we find 4 mechanisms that give rise to
  Soft X-rays: (i) The mass-loaded wind, (ii) the intermediate temperature
  interface between the hot wind and cool filaments, (iii) bow shocks, and (iv)
  interactions between bow shocks. The shell is also a major contributor to
  the soft X-ray emission, but has no associated H$\alpha$ emission.

\item The hard X-ray emission originates from gas in the starburst region.

\end{enumerate}    

The results of these simulations indicate that the host galaxy itself and the environment in
which it is situated is a major determinant in the morphology of the
outflow. The emission processes that contribute to the H$\alpha$ and soft X-ray
emission may vary from one galaxy to the next. Whether the H$\alpha$ emission
originates from photoionization or from shock-heating (or both) cannot be
determined from these simulations. However, we do find an abundance of
filamentary T $\sim$ 10$^4$ K gas that has been accelerated into the outflow, forming a
biconical shaped region that is commonly observed in starburst winds. The source
of the soft X-ray emission is also likely to depend upon the environment of
the host galaxy. In the case of M82 it is plausible that the interaction of
the wind with the surrounding HI clouds is also a contributor to the soft
X-ray emission, in addition to the processes mentioned above.

The observed spatial relationship between the H$\alpha$ and soft X-ray emitting gas can
be explained when considering emission processes interior to the wind, such as 
bow-shocks and the mass-loaded component of the wind.
In addition, the presence of the strong X-ray emitting shell with no
associated H$\alpha$ emission is interesting. 
While the ultimate fate of the shell is unknown at present, this
result argues for the presence of X-ray emission more extended than
the filamentary H$\alpha$ gas. \citet{Setal2002} suggest that this emission may
be detectable in more distant starburst galaxies. 

In future work we shall investigate the evolution of a wind over a larger time
frame and spatial extent than our current study, and look at the total energy budget of the
outflow. It is also important to further test the effect of resolution on the
filaments, and in particular the associated soft X-ray emission that arises
through mixing of the hot wind and the cooler disk gas, and will be the
subject of a subsequent paper.

\acknowledgments
JBH is supported by a Federation Fellowship from the Australian Research Council. 

\clearpage

\end{document}